\documentclass[prd,twocolumn,showpacs,showkeys,footinbib]{revtex4}
\usepackage{amsmath}
\usepackage{graphicx}

\newcommand\vek[1]{{\boldsymbol{\bf #1}}}% boldface for vectors
\newcommand\he[1]{#1^{\dagger}}% hermitian conjugate
\newcommand\gr[1]{\mathrm{#1}}% notation for groups
\DeclareMathOperator{\tr}{Tr}
\newcommand\op{\mathbf\Delta} % shorthand for the order parameter
\newcommand\sumint{\hbox{$\sum$}\!\!\!\!\!\!\!\int}
\newcommand\thname{Theorem}
\newtheorem{theorem}{\thname}
\begin{document}

\title{Helical ordering in the ground state of spin-one color superconductors\\
as a consequence of parity violation}
\author{Tom\'{a}\v{s} Brauner}
\email{brauner@ujf.cas.cz}
\affiliation{Institut f\"ur Theoretische Physik, Goethe-Universit\"at,
Max-von-Laue-Stra\ss e 1, D-60438 Frankfurt am Main,
Germany\footnote{On leave from Department of Theoretical Physics,
Nuclear Physics Institute ASCR, CZ-25068 \v Re\v z, Czech Republic}}

\begin{abstract}
We investigate spin-one color superconductivity of a single quark flavor using
the Ginzburg--Landau theory. First we examine the classic analysis of Bailin
and Love and show that by restricting to the so-called inert states, it misses
the true ground state in a part of the phase diagram. This suggests the use of
the more general, noninert states in particular within three-flavor quark
matter where the color neutrality constraint imposes stress on the spin-one
pairing and may disfavor the symmetric color-spin-locked state. In the second
part of the paper we show that, in analogy to some ferromagnetic materials,
lack of space-inversion symmetry leads to a new term in the Ginzburg--Landau
functional, which favors a spatially nonuniform long-range ordering with a
spiral structure. In color superconductors, this new parity-violating term is a
tiny effect of weak-interaction physics. The modified phase diagram is
determined and the corresponding ground states for all the phases constructed.
At the end, we estimate the coefficient of the new term in the free energy
functional, and discuss its relevance for the phenomenology of dense quark
matter.
\end{abstract}

\maketitle

\section{Introduction}
It has been known for a long time that certain magnetic materials (e.g.
$\mathrm{MnSi}$ and $\mathrm{Fe}_x\mathrm{Co}_{1-x}\mathrm{Si}$) exhibit
spatially nonuniform ordering with a long-wavelength helical structure. This
phenomenon was first explained by Dzyaloshinsky \cite{Dzyaloshinsky:1958dz} and
Moriya \cite{Moriya:1960mo} fifty years ago. They pointed out that the lack of
inversion symmetry of the underlying crystal lattice allows a new term in the
Ginzburg--Landau free energy functional (hereafter referred to as the DM term),
which makes the uniform ferromagnetic ground state unstable with respect to the
formation of the helical spin density wave. A similar effect also occurs in
antiferromagnets \cite{Shraiman:1989sh}. Microscopically, the DM term results
from the spin-orbit coupling. The theory of helical magnetism was developed to
its present form two decades after the original discovery
\cite{Bak:1980ba,Nakanishi:1980na}, and is still a subject of intensive
research in the condensed-matter-physics community
\cite{Kirkpatrick:2005ki,Binz:2006bl,Binz:2006bb}.

In the present paper we show that this phenomenon has an analogy in a vastly
different branch of physics, namely in strongly-interacting cold dense quark
matter, which exhibits color superconductivity (see \cite{Alford:2007rw} for a
recent review). The necessary ingredients for the helical ordering to take
place are a vector (spin) order parameter and lack of space-inversion symmetry.
We will therefore concentrate on spin-one color superconductivity. We will show
that in this case, the parity-violating DM term is induced by weak interactions.

Due to the exchange of gluons, the quark Cooper pairs are formed predominantly
in the spin-zero color-antitriplet channel. At very high baryon density where
the quark masses can be neglected, the cold three-flavor quark matter is known
to be in the so-called color-flavor-locked phase. However, at densities
corresponding to the neutron star cores the large value of the strange quark
mass as well as the charge neutrality constraint induce a mismatch of the Fermi
levels of different quark flavors, and thus impose stress on the cross-flavor
pairing. Other forms of pairing are then likely to occur. Depending on the size
of the Fermi surface mismatch, quarks of two flavors and two colors may combine
in the so-called 2SC phase. When the mismatch is too large, only quarks of the
same flavor can pair and the spin-one pairing then remains the only possibility
\cite{Alford:2007rw}.

Originally spin-one color superconductivity was suggested and studied as a
mechanism for pairing of quarks of a single flavor
\cite{Bailin:1984bm,Schaefer:2000tw,Hosek:2000fn,Alford:2002rz}, or a single
color \cite{Buballa:2002wy}, left over from the 2SC pairing. The classification
and physical properties of several spin-one color-superconducting phases were
worked out in a series of papers by Schmitt \emph{et al.}
\cite{Schmitt:2002sc,Schmitt:2003xq,Schmitt:2003aa,Schmitt:2004hg,
Schmitt:2004et}. Possible impacts of spin-one color superconductivity on the
phenomenology of compact stars were studied in Refs.
\cite{Aguilera:2005tg,Aguilera:2006cj,Aguilera:2006xv,Schmitt:2005ee,
Schmitt:2005er,Schmitt:2005wg,Sa'd:2006qv,Blaschke:2008br}. A different
approach to spin-one color superconductivity, based on the Schwinger--Dyson
equations, was investigated in \cite{Marhauser:2006hy}.

The weak-coupling quantum chromodynamics (QCD) calculations at asymptotically
high density show that the ground state of the single-flavor quark matter is
the color-spin-locked (CSL) phase. However, when spin-one color
superconductivity is considered within three-flavor quark matter (e.g. as a
complement to the primary 2SC pairing), the requirement of overall color
neutrality may favor other patterns of spin-one pairing \cite{Alford:2005yy}. We
will therefore in the main body of this paper perform a phenomenological
analysis based on the Ginzburg--Landau (GL) theory, treating the coefficients in
the free energy as unknown parameters.

The plan of the paper is as follows. We start in Sec. \ref{Sec:classification}
by investigating the possible forms that the order parameter in a single-flavor
spin-one color superconductor may take. Such a classification was already done
by Schmitt \cite{Schmitt:2004et} for the sake of weak-coupling QCD calculations.
However, while in QCD the order parameter is provided by a nontrivial function
of momentum near the Fermi surface---the gap function, in the GL description we
deal with a single (local) order parameter. This greatly simplifies the
analysis, allowing us to carry out a complete classification of the possible
symmetry-breaking patterns.

In Sec. \ref{Sec:GL_without_DM} we review the GL theory for spin-one color
superconductivity without the parity-violating DM term. This was already
developed by Bailin and Love more than twenty years ago \cite{Bailin:1984bm}.
Nevertheless, they restricted their attention to the so-called inert states,
proposing that one with the lowest free energy for the ground state. Here we
derive a set of inequalities between the independent quartic terms in the GL
functional which allow us to determine the unique absolute minimum of the free
energy. We thus show that there is a sector in the phase diagram where the true
ground state is actually noninert.

In Sec. \ref{Sec:GL_with_DM} we finally introduce the parity-violating DM term
into the GL free energy functional. Based on Ref. \cite{Binz:2006bb} we first
work out the (slightly generalized) theory of the helical spin density wave in
the ferromagnets. With all the necessary formalism ready, we then construct the
corresponding helical states for the spin-one color-superconducting phases, and
determine the modified phase diagram.

In Sec. \ref{Sec:microderivation} we demonstrate how the DM term arises from
weak-interaction physics. We estimate the corresponding coefficient in the
weak-coupling limit. In Sec. \ref{Sec:phenomenology} we then discuss the
possible relevance of this effect for the phenomenology of dense quark matter.
In Sec. \ref{Sec:conclusions} we summarize and conclude.

It should be noted that while some calculations such as the derivation of the
Ginzburg--Landau free energy in the high-density, weak-coupling approximation
are standard and the details may thus certainly be omitted, the algebraic
analysis presented in Sections \ref{Sec:classification},
\ref{Sec:GL_without_DM}, and \ref{Sec:GL_with_DM} is particular to spin-one
color superconductivity. Even though all derivations are based on elementary
linear algebra and require no lengthy computations, we choose to provide most
details since they cannot be found elsewhere. For reader's convenience, we
formulate some purely mathematical auxiliary material in the form of simple
theorems and defer the full proofs to the appendices.

\section{Classification of order parameters}
\label{Sec:classification} In this section we will investigate the possible
symmetry-breaking patterns in a single-flavor spin-one color superconductor. The
order parameter transforms as an antitriplet under $\gr{SU(3)}$ color
transformations, as a vector under spatial $\gr{SO(3)}$ rotations, and carries
charge of the baryon number $\gr{U(1)}$ group. It can be represented by a
complex $3\times3$ matrix, $\op_{ai}$, which transforms as \cite{Schmitt:2004et}
\begin{equation}
\op\to U\op R,
\label{transfo_rule}
\end{equation}
where $U\in\gr{SU(3)\times U(1)\equiv U(3)_L}$ and $R\in\gr{SO(3)_R}$. The
indices L and R denote the ``left'' and ``right'' symmetry groups, acting on
the order parameter. This symmetry structure is similar to that of the
superfluid Helium $3$ \cite{Vollhardt:1990vw}. However, since the symmetry
group of the spin-one color superconductors is larger than that of the
superfluid Helium [which has another $\gr{SO(3)}$ instead of the $\gr{SU(3)}$],
the classification will be somewhat simpler in the present case.

In the following it will be helpful to consider also transformations from the
``diagonal'' subgroup, $\gr{SO(3)_V}$,
\begin{equation}
\op\to R^T\op R.
\label{transfo_diag}
\end{equation}
The classification of the possible inequivalent forms of the order parameter
will be based on the following two claims which are proved in Appendix
\ref{App:OP}.
\begin{theorem}
\label{Thm:order_parameter} By a suitable symmetry transformation, the order
parameter can always be brought in the form,
\begin{equation}
\op=\begin{pmatrix}
\Delta_1 & ia_3 & -ia_2 \\
-ia_3 & \Delta_2 & ia_1 \\
ia_2 & -ia_1 & \Delta_3
\end{pmatrix}
\label{op_special}
\end{equation}
(with real parameters $\Delta_i,a_i$), being a Hermitian, positive-semidefinite
matrix.
\end{theorem}
\begin{theorem}
\label{Thm:sep_invar} Let the order parameter have the form \eqref{op_special}
and $U\in\gr{U(3)_L}$, $R\in\gr{SO(3)_V}$. Then
\begin{equation}
UR^T\op R=\op
\label{op_sep_invar}
\end{equation}
if and only if $U\op=\op$ and $R^T\op R=\op$.
\end{theorem}

Eq. \eqref{op_special} represents the simplest form to which the order
parameter can in general be cast. As could have been expected, it contains six
independent parameters: A complex $3\times3$ matrix has altogether $18$ real
parameters, $12$ of which can be fixed by a transformation from the
$12$-parametric symmetry group, $\gr{G\equiv U(3)_L\times SO(3)_R}$. We will
classify all special forms of the order parameter which leave some
\emph{continuous} subgroup of $\gr{G}$ unbroken. The analysis is greatly
simplified by {\thname} \ref{Thm:sep_invar} which ensures that one can
separately investigate invariance under left unitary transformations from
$\gr{U(3)_L}$, and diagonal orthogonal rotations from $\gr{SO(3)_V}$. [The
transformed order parameter \eqref{transfo_rule} can always be written as on
the left-hand side of Eq. \eqref{op_sep_invar} by the substitution $U\to UR^T$,
which is just another matrix from $\gr{U(3)_L}$.] There is no nontrivial
unbroken combination of transformations from the two groups. Therefore, we just
need to classify the unbroken subgroups of $\gr{U(3)_L}$ and $\gr{SO(3)_V}$.

As follows from the polar decomposition, {\thname} \ref{Thm:polar_decomp},
given in Appendix \ref{App:OP}, $\gr{U(3)_L}$ has a nontrivial unbroken
subgroup if and only if the matrix $\op$ has zero modes. Specifically, the
unbroken subgroup will be $\gr{U(\mathit{n})_L}$ where $n$ is the number of
zero modes of $\op$.

The possible unbroken subgroups of $\gr{SO(3)_V}$ may be found by elementary
geometry. The Hermitian order parameter $\op$ is written as a sum of its real
symmetric and imaginary antisymmetric parts, $\op=S+iA$. These obviously
transform separately under the diagonal subgroup $\gr{SO(3)_V}$. Moreover, the
antisymmetric part is parameterized as $A_{ij}=\epsilon_{ijk}a_k$ so that the
three components $a_i$ transform as a vector, $\vec a$. The (real) symmetric
matrix $S$ may be viewed as defining a quadratic surface with principal values
$\Delta_1,\Delta_2,\Delta_3$. Apparently, it possesses the same symmetry as $A$
if and only if this quadratic surface is axially symmetric, with the axis given
by $\vec a$. This constrains $S$ to be of the form
$$
S_{ij}=\alpha\delta_{ij}+\beta a_ia_j,
$$
or equivalently a linear combination of identity and the projector to the plane
perpendicular to $\vec a$, $\mathcal P_{ij}=\delta_{ij}-\frac{a_ia_j}{||\vec
a||^2}$. Since $S$ is actually diagonal, obviously at most one component of
$\vec a$ may be nonzero (or $\beta=0$) in order to preserve a continuous
subgroup of $\gr{SO(3)_V}$.

By combining the conditions for invariance under the left unitary and the
diagonal orthogonal transformations, one arrives at the classification
summarized in Fig. \ref{Fig:classification}. The logic of the table is simple.
First four rows display phases with nonzero $\vec a$ in order of increasing
unbroken symmetry. The next two rows show the CSL and polar phases, whose order
parameter is diagonal. Then come the phases $\gr{N_1,N_2}$ that completely
break the $\gr{SO(3)_V}$. These cannot be easily cast in the form
\eqref{op_special} and we thus display them in such a way as to manifest the
number of zero modes. Finally, the axial and planar states, showed under the
double line, are just special cases of the oblate and cylindrical ones, already
included in the table. They are distinguished by an unbroken discrete symmetry,
generated by the permutation matrix
$$
P=\begin{pmatrix}
0 & 1 & 0\\
1 & 0 & 0\\
0 & 0 & -1
\end{pmatrix},
$$
that is, a rotation by $\pi$ about the line $x=y,z=0$. We do not have the ambition to
extend the above analysis to unbroken \emph{discrete} symmetries. These two special cases
are mentioned explicitly because they will later turn out to occupy a part of the phase
diagram.

Four of the indicated phases are ``inert'', i.e., their order parameter is
fixed up to a symmetry transformation and an overall normalization. These are
the A, CSL, planar, and polar phases. Note that some of the states that are
distinct according to Ref. \cite{Schmitt:2004et}, are classified as equivalent
here. This is because (as already remarked above) we treat the matrix elements
of $\op$ as pure numbers, not functions of momentum.

A remark about the nomenclature is in order here. The names of the four inert
phases are standard in literature on spin-one color superconductivity. The terms
``oblate'' \cite{Vollhardt:1990vw} and ``$\varepsilon$'' \cite{Barton:1974ae}
have been taken over from literature on superfluid Helium 3. The remaining four
names are new. The ``N'' states are labeled by the degeneracy of the zero
eigenvalue of $\op$. In the ``cylindrical'' phase, the quadratic form $S$
defines a cylinder with the vector $\vec a$ pointing along its axis. Finally,
the ``axial'' state may be thought of as deformed CSL with just axial symmetry.
As we will see in Sec. \ref{Subsubsec:deformedCSL}, this exactly happens to the
CSL phase upon switching on the DM term \footnote{Note that in condensed-matter
physics, the term ``axial'' is sometimes used to refer to the A phase
\cite{Barton:1974ae}. In the context of the spin-one color superconductivity,
the term A phase is used exclusively so that no confusion can arise.}.
\begin{figure*}
\setlength{\tabcolsep}{0.5em}
\begin{tabular}{|c|c|c|c|}
\hline
order parameter & unbroken symmetry & name & name according to
\cite{Schmitt:2004et}\\
\hline\hline
$\begin{pmatrix}
\Delta_1 & +ia & 0\\
-ia & \Delta_1 & 0\\
0 & 0 & \Delta_2
\end{pmatrix}$
& $\gr{SO(2)_V}$ & oblate & ---\\
\hline
$\begin{pmatrix}
\Delta & +ia & 0\\
-ia & \Delta & 0\\
0 & 0 & 0
\end{pmatrix}$
& $\gr{SO(2)_V\times U(1)_L}$ & cylindrical & ---\\
\hline
$\begin{pmatrix}
\Delta_1 & +i\Delta_1 & 0\\
-i\Delta_1 & \Delta_1 & 0\\
0 & 0 & \Delta_2
\end{pmatrix}$
& $\gr{SO(2)_V\times U(1)_L}$ & $\varepsilon$ & $\gr{P_7,P_8}$\\
\hline
$\begin{pmatrix}
1 & +i & 0\\
-i & 1 & 0\\
0 & 0 & 0
\end{pmatrix}$
& $\gr{SU(2)_L\times SO(2)_V\times U(1)_L}$ & A & $\gr{A,P_3,P_5}$\\
\hline
$\begin{pmatrix}
1 & 0 & 0\\
0 & 1 & 0\\
0 & 0 & 1
\end{pmatrix}$
& $\gr{SO(3)_V}$ & CSL & CSL\\
\hline
$\begin{pmatrix}
0 & 0 & 0\\
0 & 0 & 0\\
0 & 0 & 1
\end{pmatrix}$
& $\gr{SU(2)_L\times SO(2)_R\times U(1)_L}$ & polar & $\gr{polar,P_2,P_6}$\\
\hline
$\begin{pmatrix}
0 & 0 & 0\\
z_1 & z_2 & z_3\\
z_4 & z_5 & z_6
\end{pmatrix}$
& $\gr{U(1)_L}$ & $\gr{N_1}$ & $\gr{P_4}$\\
\hline
$\begin{pmatrix}
0 & 0 & 0\\
0 & 0 & 0\\
z_1 & z_2 & z_3
\end{pmatrix}$
& $\gr{SU(2)_L\times U(1)_L}$ & $\gr{N_2}$ & $\gr{P_1}$\\
\hline\hline
$\begin{pmatrix}
\Delta_1 & 0 & 0\\
0 & \Delta_1 & 0\\
0 & 0 & \Delta_2
\end{pmatrix}$
& $\gr{SO(2)_V}$ & axial & ---\\
\hline
$\begin{pmatrix}
1 & 0 & 0\\
0 & 1 & 0\\
0 & 0 & 0
\end{pmatrix}$
& $\gr{SO(2)_V\times U(1)_L}$ & planar & planar\\
\hline
\end{tabular}
\caption{Classification of all order parameters that leave some continuous
subgroup of $\gr{G}$ unbroken. The parameters $z_i$ are complex, while all
other parameters are real, in accordance with Eq. \eqref{op_special}. For
convenience, we explicitly distinguish the $\gr{SO(2)}$ group of real rotations
from the $\gr{U(1)}$ group of phase transformations, even though the two are
actually isomorphic. The first eight rows represent all possible phases
distinguished by the unbroken continuous symmetry. The last two rows---the
axial and planar states---are merely special cases of the oblate and
cylindrical ones which have an additional discrete symmetry.}
\label{Fig:classification}
\end{figure*}

\section{Ginzburg--Landau theory without DM term}
\label{Sec:GL_without_DM} After working out the classification of all
inequivalent forms of the order parameter, we now investigate using the
Ginzburg--Landau theory, which of the states is actually energetically
preferred. Our analysis will be similar to that of Baym and Iida for spin-zero
color superconductivity \cite{Iida:2000ha}. So far, we do not include the DM
term. In the following one should always keep in mind that the GL theory is
strictly speaking only valid near the critical temperature. For the sake of
brevity, we often use the term ``ground state'' where we mean the state
minimizing the free energy, i.e., thermodynamic equilibrium. It is only in Sec.
\ref{Sec:phenomenology} that we make some speculations concerning the physics
far from the critical point, at low temperatures.

It will sometimes be convenient to use a different notation for the order
parameter, in particular to treat the matrix $\op_{ai}$ as a collection of
three complex vectors, $\vec\phi_a$, one for each anticolor $a$. The GL free
energy density up to fourth order in $\op$ then reads
\begin{multline}
\mathcal
F[\vec\phi_a]=a_1\nabla_i\he{\vec\phi}_a\cdot\nabla_i\vec\phi_a+a_2(\vec\nabla\cdot\he{\vec\phi}_a)
(\vec\nabla\cdot\vec\phi_a)\\
+b\he{\vec\phi}_a\cdot\vec\phi_a+ d_1\mathcal A+d_2\mathcal B+d_3\mathcal C.
\label{GL_without_DM}
\end{multline}
For the time being, the parameters $a_1,a_2,b,d_1,d_2,d_3$ are treated as free,
constrained only by the requirement of boundedness of the free energy from
below. There are three independent quartic $\gr{G}$-invariant terms that we
denote as $\mathcal A$, $\mathcal B$, and $\mathcal C$. Their explicit forms in
both notations for the order parameter are summarized in Fig.
\ref{Fig:invariants}.
\begin{figure}
\begin{tabular}{|c|c|c|}
\hline
name & vector expression & matrix expression\\
\hline\hline
$\mathcal A$ & $(\he{\vec\phi}_a\cdot\vec\phi_a)^2$ & $[\tr(\op\he\op)]^2$\\
\hline
$\mathcal B$ & $|\he{\vec\phi}_a\cdot\vec\phi_b|^2$ & $\tr(\op\he\op\op\he\op)$\\
\hline
$\mathcal C$ & $|\vec\phi_a\cdot\vec\phi_b|^2$ & $\tr[\op\op^T\he{(\op\op^T)}]$\\
\hline
\end{tabular}
\caption{Independent quartic invariants and their expression using the matrix $\op$
as well as the complex vectors $\vec\phi_a$.}
\label{Fig:invariants}
\end{figure}

\subsection{Ground state}
\label{Subsec:ground_state} Since the gradient part of the free energy is
required to be bounded from below, the ground state is apparently a uniform
field configuration that minimizes the static part of Eq.
\eqref{GL_without_DM}. As usual, the ``mass term'' $b$ changes sign at the
critical temperature. In the following we will assume that it is negative
(i.e., we are in the superfluid phase) so that the free energy has a nontrivial
minimum. We will use the invariant $\mathcal A$ to measure the size of the
condensate; actually, it is a squared norm of $\op$ in the sense that will be
specified later. The values of the other invariants $\mathcal B,\mathcal C$
then measure the orientation of the order parameter in the color and spin
space, and we will draw the phase diagram in the two-dimensional space of the
parameters $d_2,d_3$.

Trying to determine the ground state by a direct solution of the gap equation
would be just hopeless. First, the gap equation is a coupled set of equations
for the six independent parameters in $\op$. Second, even if we somehow managed
to solve it, we could at best show that the solution is a local minimum of the
free energy. Instead, we derive a set of inequalities between the invariants
$\mathcal A,\mathcal B,\mathcal C$ that allow us to uniquely determine the
absolute minimum of the free energy. We again formulate these inequalities as
simple theorems whose proof is given in detail in Appendix
\ref{App:inequalities}. (It is understood that all the following claims about
the order parameter hold up to a symmetry transformation.)
\begin{theorem}
\label{Thm:veta1} The invariants $\mathcal A,\mathcal B$ satisfy the following
inequalities,
\begin{equation}
\frac13\mathcal A\leq\mathcal B\leq\mathcal A. \label{veta1}
\end{equation}
The first inequality is saturated (i.e., an equality holds) if and only if the order
parameter is of the CSL type. The second inequality is saturated if and only if the matrix
$\op$ has rank one.
\end{theorem}
\begin{theorem}
\label{Thm:veta2} The invariants $\mathcal B,\mathcal C$ satisfy the following
inequalities,
\begin{equation}
0\leq\mathcal C\leq\mathcal B. \label{veta2}
\end{equation}
The first inequality is saturated if and only if the order parameter is of the A type. The
second inequality is saturated if and only if the order parameter is real.
\end{theorem}
\begin{theorem}
\label{Thm:veta3} The invariants $\mathcal A,\mathcal B,\mathcal C$ satisfy the
following inequality,
\begin{equation}
\frac23\mathcal A\leq\mathcal B+\mathcal C. \label{veta3}
\end{equation}
The inequality is saturated if and only if the order parameter is of the oblate type with
$\Delta_2=\sqrt{\Delta_1^2+a^2}$.
\end{theorem}
\begin{theorem}
\label{Thm:veta5} Let $\mathcal C\leq\frac19\mathcal A$. Then the invariants
$\mathcal A,\mathcal B,\mathcal C$ satisfy the following inequality,
\begin{equation}
\sqrt{\mathcal A}\leq\sqrt{\mathcal C}+\sqrt{\mathcal B-\mathcal C}.
\label{veta5}
\end{equation}
The inequality is saturated if and only if the order parameter is of the
$\varepsilon$ type.
\end{theorem}

With this set of inequalities at hand, at is straightforward to determine the
phase diagram in the $(d_2,d_3)$ plane. We do so by finding a lower bound on
the free energy and showing that this bound is saturated by a particular type
of the order parameter. In all cases, the free energy density can be, after
fixing the orientation of the condensate, written as
$$
\mathcal F=b\sqrt{\mathcal A}+\bar d\mathcal A,
$$
where $\bar d$ is an effective quartic coupling, specific for the given phase. The ground
state condensate and free energy are then given by
\begin{equation}
\sqrt{\mathcal A_{\text{min}}}=-\frac{b}{2\bar d},\quad \mathcal
F_{\text{min}}=-\frac{b^2}{4\bar d}.
\label{Fgrst}
\end{equation}
\begin{itemize}
\item $d_2+d_3>0,d_2>d_3$. Using {\thname}s \ref{Thm:veta2} and
\ref{Thm:veta3}, we get
\begin{multline*}
d_2\mathcal B+d_3\mathcal C=\frac12(d_2+d_3)(\mathcal B+\mathcal C)\\
+\frac12(d_2-d_3)(\mathcal B-\mathcal C)\geq\frac13(d_2+d_3)\mathcal A.
\end{multline*}
To saturate the inequality, we should have simultaneously $\mathcal B+\mathcal
C=\frac23\mathcal A$ and $\mathcal B=\mathcal C$, i.e., $\mathcal
B=\frac13\mathcal A$. By {\thname} \ref{Thm:veta1}, this is only satisfied by
the \emph{CSL} state. The magnitude of the condensate and minimum free energy
density are given by \eqref{Fgrst} with $\bar
d_{\text{CSL}}=d_1+\frac{d_2+d_3}3$.

\item $d_2+d_3<0,d_3<0$. In this case, {\thname}s \ref{Thm:veta1} and
\ref{Thm:veta2} yield
$$
d_2\mathcal B+d_3\mathcal C\geq(d_2+d_3)\mathcal B\geq(d_2+d_3)\mathcal A.
$$
The order parameter that saturates this bound should be real and have rank one, which is
precisely the \emph{polar} phase, with condensate and free energy determined by $\bar
d_{\text{polar}}=d_1+d_2+d_3$.

\item $d_2<0,d_3>0$. Now we estimate the free energy using {\thname}s \ref{Thm:veta1} and
\ref{Thm:veta2} as
$$
d_2\mathcal B+d_3\mathcal C\geq d_2\mathcal A.
$$
The ground state is the \emph{A} phase. The effective quartic coupling in this case reads
$\bar d_{\text{A}}=d_1+d_2$.

\item $d_3>d_2>0$. This case is subtle; it is here that none of the inert phases
provides the absolute minimum of the free energy. Let us assume that $\mathcal
C\leq\frac19\mathcal A$. We then use {\thname} \ref{Thm:veta5} and the Cauchy
inequality \eqref{cauchy} with $u_1=\sqrt{d_2(\mathcal B-\mathcal
C)},u_2=\sqrt{(d_2+d_3)\mathcal C},v_1=1/\sqrt{d_2},v_2=1/\sqrt{d_2+d_3}$, to
obtain
\begin{multline*}
d_2\mathcal B+d_3\mathcal C=d_2(\mathcal B-\mathcal C)+(d_2+d_3)\mathcal C=u_1^2+u_2^2\geq\\
\geq\frac{(u_1v_1+u_2v_2)^2}{v_1^2+v_2^2}= \frac{(\sqrt{\mathcal B-\mathcal
C}+\sqrt{\mathcal C})^2}{\frac1{d_2}+\frac1{d_2+d_3}}\geq
\frac{d_2(d_2+d_3)}{2d_2+d_3}\mathcal A.
\end{multline*}
The inequality is saturated if and only if the order parameter is of the
\emph{$\varepsilon$} type, and vectors $\vec u$ and $\vec v$ are collinear.
This fixes the order parameter to be
\begin{equation}
\begin{split}
\op&=\sqrt{\mathcal A_{\text{min}}}\begin{pmatrix}
\alpha & i\alpha & 0\\
-i\alpha & \alpha & 0\\
0 & 0 & \beta
\end{pmatrix},\\
\alpha&=\frac12\sqrt{\frac{d_2+d_3}{2d_2+d_3}},\quad
\beta=\sqrt{\frac{d_2}{2d_2+d_3}}.
\end{split}
\label{oblate}
\end{equation}
The effective quartic coupling in this case is $\bar
d_\varepsilon=d_1+\frac{d_2(d_2+d_3)}{2d_2+d_3}$. One easily checks that the
initial assumption $\mathcal C\leq\frac19\mathcal A$ is fulfilled for
considered values of $d_2,d_3$.
\end{itemize}

\subsection{Phase diagram}
\label{Subsec:phase_diagram_without} The calculation of the ground state for
different relative values of $d_2,d_3$ given in the previous subsection is
straightforward, but may not be entirely transparent. Therefore, we complement
it here by an elegant and powerful geometric picture, first developed by Kim
and Frautschi \cite{Kim:1981xu,Frautschi:1981jh} to analyze complicated Higgs
potentials in models of grand unification. It will not only confirm our
previous conclusions about the ground state, but also illuminate the nature of
the inert and noninert states and the phase transitions between them.

The idea is as follows. The quartic part of the free energy can be thought of
as depending on the squared norm of the condensate, $\mathcal A$, and two
dimensionless quantities, $\lambda_2=\mathcal B/\mathcal A$ and
$\lambda_3=\mathcal C/\mathcal A$. These specify the orientation of the
condensate in the color and spin space. For a uniform field configuration, the
free energy density thus becomes
\begin{equation}
\mathcal F_{\text{stat}}=b\sqrt{\mathcal
A}+(d_1+d_2\lambda_2+d_3\lambda_3)\mathcal A. \label{Flambda}
\end{equation}
The inequalities derived above show that the quantities $\lambda_2,\lambda_3$
cannot acquire arbitrary values. Instead, their values for all nonzero
$3\times3$ matrices will span some domain in the $(\lambda_2,\lambda_3)$ plane,
which we will refer to as the target space, in order to distinguish it from the
parameter space of $d_2,d_3$. The shape of the target space is a property of
the algebra of $3\times3$ matrices (and the symmetry group $\gr G$), and is
independent of the couplings $d_2,d_3$.

The absolute minimum of the free energy \eqref{Flambda} can now be found by a
consecutive minimization with respect to the ``angles'' $\lambda_2,\lambda_3$,
and then the ``modulus'' $\sqrt{\mathcal A}$. But since the free energy
\eqref{Flambda} is linear in $\lambda_2,\lambda_3$, the minimum will simply lie
somewhere on the boundary of the target space. Which point of the boundary will
realize the ground state, depends on the coefficients $d_2,d_3$. For fixed
values of $\mathcal F_{\text{stat}}$ and $\sqrt{\mathcal A}$, Eq.
\eqref{Flambda} defines a straight line in the $(\lambda_2,\lambda_3)$ plane.
For a too small value of $\mathcal F_{\text{stat}}$, this line will not
intersect the target space, i.e., there is no state with the desired value of
the free energy. As $\mathcal F_{\text{stat}}$ increases, the straight line
will shift parallel until for some $\mathcal F_{\text{min}}$, it will for the
first time touch the target space. The point of touch will then define the
ground state.

\begin{figure}
\includegraphics[scale=0.45]{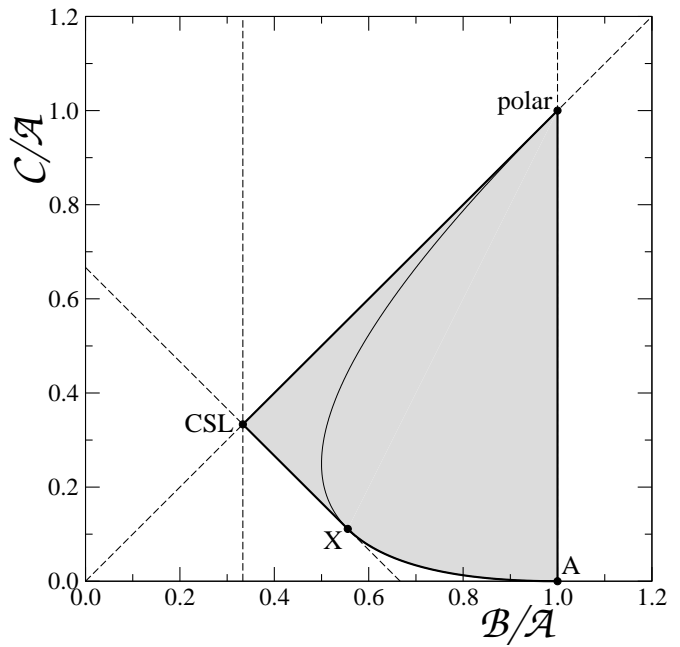}
\caption{Target space in the $(\lambda_2,\lambda_3)$ plane (shaded). The
various bounding curves are determined by Eqs. \eqref{veta1}--\eqref{veta5}.
The curve connecting the points ``A'', ``X'', and ``polar'' is defined by
$\sqrt{\mathcal C}+\sqrt{\mathcal B-\mathcal C}=\sqrt{\mathcal A}$. For given
$d_2,d_3$ the structure of the ground state is determined by that point on the
boundary of the target space which minimizes the expression
$d_2\lambda_2+d_3\lambda_3$, i.e., by the nearest point of the target space
when looking at it (in the plane) in the direction of the vector $(d_2,d_3)$.}
\label{Fig:Kim_plot}
\end{figure}
In the case of the spin-one color superconductor, the explicit form of the
target space is plotted in Fig. \ref{Fig:Kim_plot}. We emphasize once again
that it is the points at its boundary that will appear in the phase diagram as
ground states for some particular combination of $d_2,d_3$. The states that
correspond to the corners of the target space therefore play a distinguished
role. These are the three inert states, A, CSL, and polar. Since they are
inert, they are represented by a single point in the target space. A noninert
state of a particular type will occupy some nontrivial domain, depending on its
number of free parameters.

Let us now be more specific about the boundary of the target space. The second
inequality in Eq. \eqref{veta1} defines the edge connecting the ``A'' and
``polar'' corners, which thus involves all rank-one order parameters. The
polar--CSL edge follows from the second inequality in Eq. \eqref{veta2} and
consists of matrices $\op$ that can be made real by a symmetry transformation.
The CSL--X edge comes from Eq. \eqref{veta3}. It is occupied by matrices of the
oblate type with $\Delta_2=\sqrt{\Delta_1^2+a^2}$. Finally, the curved segment
X--A is a consequence of {\thname} \ref{Thm:veta5} and is realized by the
matrices of the $\varepsilon$ type. This completes the picture of the target
space and elucidates the significance of the various inequalities.

\begin{figure}
\includegraphics[scale=0.45]{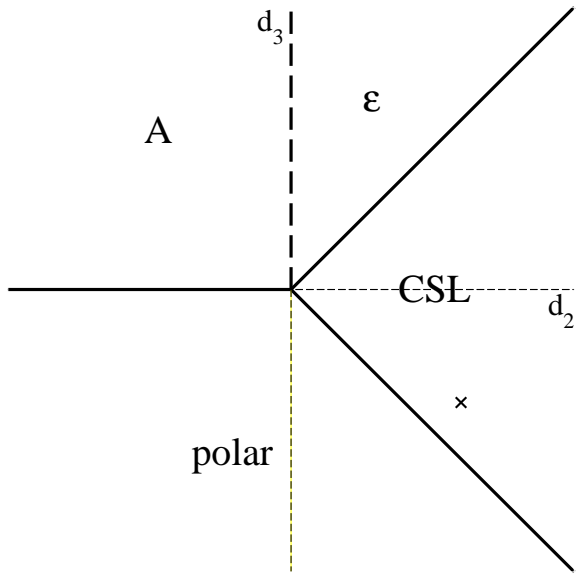}
\caption{Phase diagram of the spin-one color superconductor in the $(d_2,d_3)$
plane. The solid and dashed thick lines denote first- and second-order phase
transitions, respectively. The phase boundaries are defined by straight radial
lines at angles $\frac\pi4,\frac\pi2,\pi,\frac{7\pi}4$ with respect to the
$d_2$ axis. The cross indicates the weak-coupling prediction, see Eq.
\eqref{GL_micro_without}.} \label{Fig:phase_diagram_without}
\end{figure}

The analysis carried out in Sec. \ref{Subsec:ground_state} yields the phase
diagram, displayed in Fig. \ref{Fig:phase_diagram_without}. Note that all phase
transitions but the one between the A and $\varepsilon$ phases are first order.
This is also easily seen using the Kim--Frautschi plot of the target space (see
Fig. \ref{Fig:Kim_plot}). A straight segment of the boundary connecting two
corners such as polar and CSL causes the ground state to change abruptly at an
infinitesimal change of the slope of the straight line \eqref{Flambda} (i.e.,
the ratio of $d_2$ and $d_3$). On the other hand, as we move through the
$\varepsilon$ phase towards the transition line to the A phase, the order
parameter runs along the curved segment X--A until it eventually continuously
enters the A phase. Moreover, we can easily check explicitly using Eq.
\eqref{oblate} that as $d_2\to0+$, the $\varepsilon$ state continuously goes to
the A state.

The Kim--Frautschi plot also tells us which types of order parameters may coexist right at
the first-order phase transition lines. If the segment of the boundary of the target space
connecting the two competing phases were concave, we would find just these two states.
However, since all the border lines corresponding to first-order phase transitions are
straight, a much wider class of states can actually coexist. Without going into details we
just note that special relations between $d_2$ and $d_3$ which define the phase transition
lines, may bring in additional degeneracy in terms of an enhanced symmetry of the free
energy \cite{Brauner:2005di}.

Note that the phase diagram in Fig. \ref{Fig:phase_diagram_without} does not
depend on the coefficient $d_1$. The only way it affects the problem is
indirectly, by the requirement of boundedness of the free energy from below. In
other words, a particular value of $d_1$ will determine a region in the
$(d_2,d_3)$ plane which is physically allowed.

Finally, a comparison with the phase diagram calculated in Ref.
\cite{Bailin:1984bm} [and plotted in the $(d_2,d_2+d_3)$ plane] shows that our
results agree with the exception of the $\varepsilon$ region, which was missed
in \cite{Bailin:1984bm}. The strategy used here, based on lower estimates of
the free energy together with the conditions for their saturation, ensures that
we have really found the absolute minimum of the free energy. Apart from the
phase transition lines, it is unique up to a symmetry transformation.

\section{Ginzburg--Landau theory with DM term}
\label{Sec:GL_with_DM} We are now ready to analyze the GL theory for spin-one
color superconductivity including the parity-violating DM term. However, since
the construction of the helix-ordered state for some of the
color-superconducting phases is a bit involved, we prefer to illustrate
the idea and develop the argument on a simple toy example.

\subsection{Toy model: Complex ferromagnet}
\label{Subsec:complex_ferro} Following closely Ref. \cite{Binz:2006bb}, we
consider the GL free energy density functional of the form \footnote{In a
ferromagnet the order parameter $\vec M$ would, of course, be real. Also, the
$a_2$ term would be absent owing to the fact that there are no magnetic
charges. This slight generalization of the GL functional of Ref.
\cite{Binz:2006bb} allows us to work out an analysis which will later apply
almost without change to spin-one color superconductivity.}
\begin{multline}
\mathcal F[\vec M]=a_1\nabla_i\he{\vec M}\cdot\nabla_i\vec
M+a_2(\vec\nabla\cdot\he{\vec M})
(\vec\nabla\cdot\vec M)\\
+b\he{\vec M}\cdot\vec M+ c\he{\vec M}\cdot(\vec\nabla\times\vec M)+d(\he{\vec
M}\cdot\vec M)^2, \label{GL_Binz}
\end{multline}
where the term proportional to $c$ is the DM term. (Note that it is real up to
a total derivative.) Without lack of generality we will assume that $c>0$.

First of all we would like to stress that since the DM term will make the
ground state configuration nonuniform, we are not a priori allowed to simply
minimize the static part of the free energy to determine the magnitude of the
order parameter $\vec M$ (which we shall in this section refer to as the
magnetization). Instead, we will rely on an estimate of the free energy, this
time including the space dependence of the order parameter. Assuming the space
has finite volume $\Omega$ with periodic boundary conditions, we expand the
magnetization in Fourier modes,
$$
\vec M(\vek x)=\sum_{\vek k}\vec m_{\vek k}e^{i\vek k\cdot\vek x}.
$$
Using the integral form of the Cauchy inequality, we may estimate the quartic part of the
free energy,
\begin{multline*}
\int d^3\vek x\,(\he{\vec M}\cdot\vec M)^2\geq\frac1\Omega\left(\int d^3\vek
x\,\he{\vec M}\cdot\vec M\right)^2\\
=\Omega\biggl(\sum_{\vek k}|\vec m_{\vek k}|^2\biggr)^2\equiv\Omega\mathcal
M^2.
\end{multline*}
The inequality is saturated if $\he{\vec M}\cdot\vec M$ is uniform, i.e., the
magnetization has the same magnitude in the whole space.

Decomposing the Fourier mode $\vec m_{\vek k}$ into components parallel
(``longitudinal'') and perpendicular (``transverse'') to the momentum $\vek k$,
$\vec m_{\vek k\parallel}$ and $\vec m_{\vek k\perp}$, and the transverse
component further into its real and imaginary parts, $\vec m_{\vek k\perp}=\vec
u_{\vek k}+i\vec v_{\vek k}$, the DM contribution to the free energy becomes
$$
\int d^3\vek x\,\he{\vec M}\cdot(\vec\nabla\times\vec M)=2\Omega\sum_{\vek
k}\vek k\cdot(\vec u_{\vek k}\times\vec v_{\vek k}).
$$
Using elementary geometry, this is estimated as
\begin{multline*}
\vek k\cdot(\vec u_{\vek k}\times\vec v_{\vek k})\geq-|\vek k|\,|\vec u_{\vek
k}\times\vec v_{\vek k}|\geq\\
\geq-|\vek k|\,|\vec u_{\vek k}|\,|\vec v_{\vek k}|\geq-|\vek k|\frac{|\vec
u_{\vek k}|^2+|\vec v_{\vek k}|^2}2=-\frac12|\vek k|\,|\vec m_{\vek k\perp}|^2.
\end{multline*}
The chain of inequalities is saturated if the real and imaginary parts of $\vec
m_{\vek k\perp}$ have the same size, are perpendicular to each other as well as
to $\vek k$, and together form a left-handed orthogonal system of vectors. (For
$c<0$, it would be right-handed.) Together with the gradient terms, the DM term
can thus be minimized as follows,
\begin{multline*}
\int d^3\vek x\,[a_1|\nabla_i\vec M|^2+a_2|\vec\nabla\cdot\vec M|^2+c\he{\vec
M}\cdot(\vec\nabla\times\vec M)]\geq\\
\geq\Omega\sum_{\vek k}\left[(a_1+a_2)|\vek k|^2|\vec m_{\vek k\parallel}|^2+
(a_1|\vek k|^2-c|\vek k|)|\vec m_{\vek k\perp}|^2\right].
\end{multline*}
Stability of the system with respect to longitudinal fluctuations requires
$a_1+a_2>0$. The longitudinal mode $\vec m_{\vek k\parallel}$ therefore always
increases the free energy. On the other hand, \emph{the DM term, being linear
in momentum, can outweigh the gradient term and make nonuniform, transverse
field configurations energetically favorable.} The minimum free energy is
achieved when only modes with $|\vek k|=|\vek k|_{\text{min}}=c/2a_1$ are
included.

Putting all the pieces together, we obtain the minimum free energy density as
\begin{equation}
\frac1\Omega\int d^3\vek x\,\mathcal F[\vec
M]\geq\left(b-\frac{c^2}{4a_1}\right)\mathcal M+d\mathcal M^2. \label{FminBinz}
\end{equation}
The state minimizing the free energy will be purely transverse so that the
$a_2$ term in Eq. \eqref{GL_Binz} actually does not play any role. To specify
the form of the ground state more concretely, recall that apart from being
composed solely of Fourier modes with $|\vek k|=|\vek k|_{\text{min}}$, it also
ought to have a spatially uniform magnitude of magnetization. The most general
state satisfying this condition has the form
\begin{equation}
\vec M=\alpha\vec m_{\vek k}e^{i\vek k\cdot\vek x}+\beta\vec m_{\vek
k}^*e^{-i\vek k\cdot\vek x},
\label{complex_helix}
\end{equation}
with fixed momentum $\vek k$ and real coefficients $\alpha,\beta$.

Two special cases deserve particular attention. First, if $\alpha$ or $\beta$ is zero, the
ground state is a \emph{single} (complex) \emph{plane wave}. Second, if we require the
order parameter to be real (which is the case of the ferromagnet as well as several of the
spin-one color-superconducting phases), we arrive at
$$
\vec M=\vec m_{\vek k}e^{i\vek k\cdot\vek x}+\vec m_{\vek k}^*e^{-i\vek
k\cdot\vek x}=2[\vec u_{\vek k}\cos(\vek k\cdot\vek x)-\vec v_{\vek k}\sin(\vek
k\cdot\vek x)],
$$
i.e., a \emph{real standing wave}. The magnetization evolves along a
\emph{right-handed helix} with the axis defined by the vector $\vek k$, and the
wavelength $\lambda=4\pi a_1/c$. This concludes the argument and reveals the
nature of the ground state induced by the DM term.

Several remarks are in order here. First, for generic (nonzero) coefficients
$\alpha,\beta$, the state \eqref{complex_helix} breaks both rotational and
translational invariance \footnote{While in ordinary ferromagnets, $\gr{SO(3)}$
rotational invariance is broken down to $\gr{SO(2)}$ rotations about the
direction of the magnetization, here the rotational invariance is broken
completely, thanks to the presence of another vector, $\vek k$. The group of
translations along $\vek k$ is broken down to its discrete subgroup, defined by
the wavelength of the helix. Note that this still allows quasiparticle
low-energy excitations with a well defined momentum.}. However, there is a
combination of a rotation about $\vek k$ and a simultaneous translation along
$\vek k$ which remains unbroken. This leads to a peculiar, strongly anisotropic
behavior of the Nambu--Goldstone mode of the broken symmetry
\cite{Kirkpatrick:2005ki}.

Second, the wavelength of the helical state is proportional to $1/c$, hence the weaker is
the DM term, the longer is the scale of the helical ordering. This will be particularly
important in the later application to spin-one color superconductivity where the
$c$-term comes from weak interactions and is thus expected to be tiny.

Third, according to Eq. \eqref{FminBinz}, the DM term effectively lowers the coefficient
$b$. Therefore, it leads to a slight increase of the critical temperature, and at a fixed
temperature to a slight increase of the magnitude of the magnetization. However, its most
notable consequence is the formation of the nonuniform field configuration.

Fourth, as long as we restrict ourselves to terms up to fourth order in $\op$
in the free energy, the ground state can be rigorously proved to be composed of
a single plane wave (and possibly the counterpropagating wave), although the
gradient terms are minimized by Fourier modes with momenta lying on a sphere of
radius $|\vek k|_{\text{min}}$.

\subsection{Spin-one color superconductor}
\label{Subsec:CSC_helix} The generalization of the GL functional
\eqref{GL_Binz} to the case of the spin-one color superconductor is
straightforward and unique,
\begin{multline}
\mathcal
F[\vec\phi_a]=a_1\nabla_i\he{\vec\phi}_a\cdot\nabla_i\vec\phi_a+a_2(\vec\nabla\cdot\he{\vec\phi}_a)
(\vec\nabla\cdot\vec\phi_a)\\
+b\he{\vec\phi}_a\cdot\vec\phi_a
+c\he{\vec\phi}_a\cdot(\vec\nabla\times\vec\phi_a)+d_1\mathcal A+d_2\mathcal
B+d_3\mathcal C.
\label{GL_with_DM}
\end{multline}
As above, we Fourier-decompose the order parameter field,
\begin{equation}
\vec\phi_a(\vek x)=\sum_{\vek k}\vec\varphi_{a\vek k}e^{i\vek k\cdot\vek x},
\label{OP_FT}
\end{equation}
and define $\mathcal M\equiv\sum_{\vek k}\he{\vec\varphi}_{a\vek
k}\cdot\vec\varphi_{a\vek k}$. Using the same argument as in the preceding
subsection, the free energy is estimated as
\begin{equation}
\frac1\Omega\int d^3\vek x\,\mathcal
F[\vec\phi_a]\geq\left(b-\frac{c^2}{4a_1}\right)\mathcal M+\bar d\mathcal M^2,
\label{FminCSC}
\end{equation}
where the effective quartic coupling $\bar d$ for the various spin-one
color-superconducting phases was defined in Sec. \ref{Subsec:ground_state}. The
absolute minimum of the free energy is achieved for purely transverse
configurations with uniform summed magnitude $\he{\vec\phi}_a\cdot\vec\phi_a$
such that \emph{for all colors} $a$, the real and imaginary parts of
$\vec\varphi_{a\vek k}$ have the same magnitude and together with $\vek k$ form
a left-handed orthogonal system of vectors, and only modes with $|\vek k|=|\vek
k|_{\text{min}}=c/2a_1$ are included. We shall now construct the helical states
for the phases that appear in the phase diagram in Fig.
\ref{Fig:phase_diagram_without}.

\subsubsection{Polar and A phases}
A single anticolor participates in the condensation. The ground state can
therefore be constructed in complete analogy with Sec.
\ref{Subsec:complex_ferro}. For the polar phase, the order parameter is real,
and we thus find a real standing wave. Choosing the coordinate basis so that
the helix points along the $z$ axis, the order parameter takes the form
$$
\op_{\text{polar}}=\sqrt{\mathcal A_{\text{min}}}\begin{pmatrix}
0 & 0 & 0\\
0 & 0 & 0\\
\cos kz & \sin kz & 0
\end{pmatrix}.
$$
The condensate magnitude and free energy are given by a modification of Eq.
\eqref{Fgrst},
\begin{equation}
\sqrt{\mathcal A_{\text{min}}}=-\frac{b(1+\xi)}{2\bar d},\quad \mathcal
F_{\text{min}}=-\frac{b^2(1+\xi)^2}{4\bar d},
\label{Fgrsthelix}
\end{equation}
where
\begin{equation}
\xi\equiv-c^2/{4a_1b},
\label{xi_definition}
\end{equation}
and as before, $\bar d_{\text{polar}}=d_1+d_2+d_3$.

In the case of the $A$ phase, the order parameter can also be cast in such a
form that only one anticolor condenses. (The matrix $\op$ is then of course no
longer Hermitian.) The nonzero vector $\vec\phi_a$ is forced to be ``maximally
complex'' in the sense that $\mathcal C=0$. This leads to the single plane wave
solution,
$$
\op_{\text{A}}=\sqrt{\frac{\mathcal A_{\text{min}}}2}\,e^{ikz}
\begin{pmatrix}
0 & 0 & 0\\
0 & 0 & 0\\
1 & -i & 0
\end{pmatrix}.
$$
The condensate magnitude and free energy follow from Eq. \eqref{Fgrsthelix} with $\bar
d_{\text{A}}=d_1+d_2$.

\subsubsection{$\varepsilon$ phase}
In this phase, the order parameter can be cast in a form that just two
anticolors condense, which amounts to adding another orthogonal real vector to
the above A-phase order parameter. However, three real vectors cannot be
simultaneously orthogonal to each other, and still orthogonal to the momentum
$\vek k$. This means that the lower bound \eqref{FminCSC} cannot be reached,
or, the gradient and static parts of the free energy cannot be separately
minimized for an order parameter of the $\varepsilon$ type.

To see what the ground state will look like, we resort to the Ginzburg--Landau
equation following from Eq. \eqref{GL_with_DM},
\begin{multline}
-a_1\vec\nabla^2\vec\phi_a-a_2\vec\nabla(\vec\nabla\cdot\vec\phi_a)
+b\vec\phi_a+c(\nabla\times\vec\phi_a)\\
+2d_1\vec\phi_a(\he{\vec\phi}_b\cdot\vec\phi_b)
+2d_2\vec\phi_b(\vec\phi_a\cdot\he{\vec\phi}_b)+
2d_3\he{\vec\phi}_b(\vec\phi_a\cdot\vec\phi_b)=0.
\label{GLeq}
\end{multline}
We are not going to solve this equation directly. After all, even if we did, we
could not prove anyway that our solution was the absolute minimum of the free
energy. Nonetheless, Eq. \eqref{GLeq} will provide us with the necessary
insight to make a heuristic guess about the form of the ground state.

The starting assumption is that the values of the order parameter $\vec\phi_a$
at different space points are connected by a symmetry transformation. This is
reasonable for otherwise the static part of the free energy density would not
be uniform; it is hard to imagine how such a configuration could be even a
local minimum of the free energy.

The DM term in the ``equation of motion'' \eqref{GLeq} forces the vector
$\vec\phi_a$ to rotate about the direction of $\vek k$. Also, this term does
not mix colors. Hence it is plausible that the transformation which connects
the values of the order parameter at different points is a pure spatial
rotation. Naturally, the axis of rotation is to be identified with the
direction of the momentum $\vek k$. We can then choose the basis in the color
space in such a way that for one anticolor, the vector $\vec\phi_a$ is
perpendicular to $\vek k$ and rotates about it (transverse mode), while the
vector of the second anticolor is aligned with $\vek k$ and static
(longitudinal mode). This is also in accord with our general discussion since a
longitudinal mode with nonzero momentum would cost energy.

With the above argument in mind, we write down the helical state of the
$\varepsilon$ type with $\vek k$ along the $z$ axis,
\begin{equation}
\op_{\text{$\varepsilon$}}=\sqrt{\mathcal A_{\text{min}}}\begin{pmatrix}
0 & 0 & 0\\
\sqrt2\alpha e^{ikz} & -i\sqrt2\alpha e^{ikz} & 0\\
0 & 0 & \beta
\end{pmatrix}
\label{oblate_ansatz}
\end{equation}
Here $k,\alpha,\beta$, and $\sqrt{\mathcal A_{\text{min}}}$ are treated as
variational parameters with the constraint $4\alpha^2+\beta^2=1$ enforced by
normalization. Minimizing the free energy within this class of variational
states, we find
\begin{align*}
4\alpha^2&=\frac{d_2+d_3+\xi(d_1+d_2+d_3)}{d_2(2+\xi)+d_3(1+\xi)},\\
\beta^2&=\frac{d_2-\xi d_1}{d_2(2+\xi)+d_3(1+\xi)},
\end{align*}
and the momentum is, as above, given by $k=c/2a_1$. Note that for $\xi=0$ these
expressions reduce to the previous result, Eq. \eqref{oblate}. The condensate
magnitude and free energy read
\begin{align*}
\sqrt{\mathcal A_{\text{min}}}&=-\frac
b2\frac{1+\xi\frac{d_2+d_3}{2d_2+d_3}}{d_1+d_2\frac{d_2+d_3}{2d_2+d_3}},\\
\mathcal
F_{\text{min}}&=-\frac{b^2}4\frac{1+2\xi\frac{d_2+d_3}{2d_2+d_3}+\xi^2\frac{d_1+d_2+d_3}{2d_2+d_3}}
{d_1+d_2\frac{d_2+d_3}{2d_2+d_3}}.
\end{align*}
While the denominators of the large fractions contain the expected effective
quartic coupling, $\bar
d_{\text{$\varepsilon$}}=d_1+d_2\frac{d_2+d_3}{2d_2+d_3}$, the numerators
differ from Eq. \eqref{Fgrsthelix}. This is because the separate minimization
of the gradient and static parts of the free energy could not be achieved so
that both the condensate and the condensation energy are actually smaller than
Eq. \eqref{Fgrsthelix} would have predicted.

Apart from the variational minimization of the free energy with the ansatz
\eqref{oblate_ansatz}, we also checked explicitly that this state with parameters fixed by
the above expressions indeed solves the GL equation \eqref{GLeq}. This provides a decent
evidence that we have found the genuine ground state.

\subsubsection{CSL phase}
\label{Subsubsec:deformedCSL} Here the situation is even more severe than for
the $\varepsilon$ phase. In that case, the order parameter contains even after
normalization a free parameter, say, the ratio $\alpha/\beta$. The effect of
the DM term in the free energy is then accounted for by a slight shift of this
parameter. However, the CSL order parameter is rigid, there are no free
parameters beyond the overall norm to adjust. As a result, the CSL state simply
turns out to be incompatible with the DM term: It is no longer a minimum of the
free energy.

To see this, let us assume [as in the discussion below Eq. \eqref{GLeq}] that the order
parameter has everywhere the CSL form with a fixed magnitude. Consequently, the
matrix $\op$ is unitary up to a real coordinate-independent factor. This dramatically
simplifies the GL equation \eqref{GLeq}, which becomes linear and separated for the
individual colors,
$$
-a_1\vec\nabla^2\vec\phi_a-a_2\vec\nabla(\vec\nabla\cdot\vec\phi_a)
+c\vec\nabla\times\vec\phi_a+b_{\text{eff}}\vec\phi_a=0,
$$
with $b_{\text{eff}}=b+2\bar d_{\text{CSL}}\sqrt{\mathcal A}$. Being linear,
this equation can be solved independently for the static and rotating modes. In
particular the existence of a static mode requires $b_{\text{eff}}=0$. This
fixes the condensate magnitude to its size without the DM term. The rotating
mode has to fulfill the condition $a_1|\vek k|^2=c|\vek k|$. Therefore, even
though there is a solution with helical structure, its energy is the same as
that of the uniform CSL condensate.

In order to resolve this problem, it is important to realize that the anticipated
existence of a helical structure in the ground state implies breaking of the rotational
symmetry at least to the group of rotations about the axis of the helix, $\vek k$. It
would be naive to expect the isotropic CSL state in such a situation.

We can take the reduced axial symmetry into account and at the same time relax
the rigidity of the order parameter by considering the more general
\emph{axial} state (see Fig. \ref{Fig:classification}). This is natural: There
is no reason why the static part of the order parameter, aligned with $\vek k$,
should have the same length as the transverse part, which is perpendicular to
$\vek k$ and rotates about it. In fact, thanks to the DM term, we should expect
the transverse part to be preferred. This argument leads us to the ansatz
\begin{equation}
\op_{\text{axial}}=\sqrt{\mathcal A_{\text{min}}}\begin{pmatrix}
\alpha\cos kz & \alpha\sin kz & 0\\
-\alpha\sin kz & \alpha\cos kz & 0\\
0 & 0 & \beta
\end{pmatrix},
\label{axial:OP}
\end{equation}
with the normalization constraint $2\alpha^2+\beta^2=1$. The variational minimization of
the free energy now results in
$$
\alpha^2=\frac{1+\xi\frac{d_1+d_2+d_3}{d_2+d_3}}{3+2\xi},\quad
\beta^2=\frac{1-2\xi\frac{d_1}{d_2+d_3}}{3+2\xi}.
$$
The condensate magnitude and free energy are given by
\begin{align*}
\sqrt{\mathcal A_{\text{min}}}&=-\frac b2\frac{3+2\xi}{3d_1+d_2+d_3},\\
\mathcal
F_{\text{min}}&=-\frac{b^2}4\frac{3+4\xi+2\xi^2\frac{d_1+d_2+d_3}{d_2+d_3}}{3d_1+d_2+d_3}.
\end{align*}
These results show that $\alpha$ is always larger than $\beta$ so that the
rotating part of the order parameter is indeed favored over the static one, as
predicted. In comparison to the CSL state this physically means that one can
gain energy by making the condensates of the three anticolors slightly
imbalanced in favor of the two anticolors that form the helical structure. This
has the amusing consequence that the ground state is no longer color-neutral.
The color-density imbalance is a sheer weak-interaction effect.

\subsubsection{Planar phase}
The magnitude $\beta$ of the static part of the axial order parameter decreases with the
sum $d_2+d_3$ until at $d_2+d_3=2\xi d_1$ it goes to zero. The axial order parameter
reduces to the planar one. Since in the phase diagram without the DM term, Fig.
\ref{Fig:phase_diagram_without}, the boundary between the CSL and polar phases occurs at
$d_2+d_3=0$, we may expect the planar phase to interpose itself between the two at nonzero
$\xi$.

The planar order parameter is explicitly expressed as
\begin{equation}
\op_{\text{planar}}=\sqrt{\frac{\mathcal A_{\text{min}}}2}\begin{pmatrix}
\cos kz & \sin kz & 0\\
-\sin kz & \cos kz & 0\\
0 & 0 & 0
\end{pmatrix},
\label{planar:OP}
\end{equation}
and its size and free energy are given by Eq. \eqref{Fgrsthelix} with $\bar
d_{\text{planar}}=d_1+\frac{d_2+d_3}2$.

\subsection{Phase diagram}
\label{Subsec:phase_diagram_with}
\begin{figure}
\includegraphics[scale=0.45]{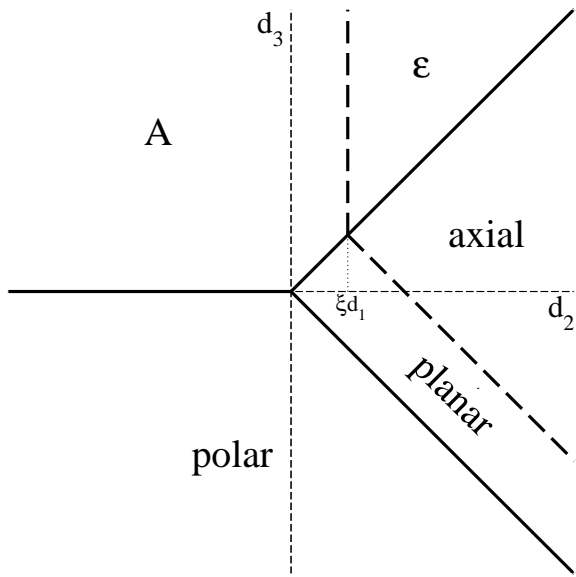}
\caption{Phase diagram of the spin-one color superconductor including the DM
term. The solid and dashed thick lines denote first- and second-order phase
transitions, respectively. The positions of the solid lines are the same as in
Fig. \ref{Fig:phase_diagram_without} and independent of the DM term, whereas
the offset of the dashed lines is proportional to the parameter $\xi$ defined
in Eq. \eqref{xi_definition}.} \label{Fig:phase_diagram_with}
\end{figure}
The phase diagram, modified by the parity-violating DM term, is displayed in
Fig. \ref{Fig:phase_diagram_with}, assuming that $d_1>0$. [For $d_1<0$, a large
part of the $(d_2,d_3)$ plane would be excluded by the requirement of the
boundedness of the free energy so that the resulting picture would not be very
interesting.] The $\varepsilon$--A and axial--planar phase transitions are of
the second order. They are both characterized by vanishing of the
respective $\beta$ parameter. The rest of the phase transitions are first order.

For the axial and $\varepsilon$ states, the minimum free energy suggested by
Eq. \eqref{FminCSC} cannot be achieved. As a result, these two phases are
pushed away from the phase diagram by the DM term. The first-order transition
lines were determined by a comparison of the free energies of the individual
phases. It was thus checked that the free energy is continuous all over the
phase diagram, which is a necessary condition for the conclusion that we have
not missed any possible intermediate phase and determined the phase diagram
correctly.

Finally, note that while the complete GL free energy \eqref{GL_with_DM} is
$\gr{SU(3)\times SO(3)\times U(1)}$-invariant, the DM term has in fact an
enhanced $\gr{SO(6)\times SO(3)}$ symmetry under which the pairs of states
A--planar and $\varepsilon$--axial are degenerate. This has in particular the
consequence that the $\varepsilon$--axial transition line in the phase diagram
(the CSL state being a special case of the axial one) does not shift when the
DM term is switched on.

\section{Microscopic derivation of DM term}
\label{Sec:microderivation} In this section, we provide a microscopic
derivation of the Ginzburg--Landau free energy. Again, a thorough analysis of
this problem was performed by Bailin and Love long time ago. Nevertheless,
since the calculation at the level of generality taken up in Ref.
\cite{Bailin:1984bm} is rather involved, we make a number of simplifying
assumptions that allow us to arrive at a simple formula in an efficient manner.

\subsection{GL functional from NJL model}
First of all, we fix the Dirac structure of the spin-one gap matrix to be
simply $\gamma_i$. Such an order parameter has positive parity and in the
ultrarelativistic limit describes purely transverse pairing of fermions, which
was shown in Ref. \cite{Schmitt:2004et} to be energetically preferred to the
longitudinal pairing \footnote{We stress that the terms transverse and
longitudinal here, commonly used in literature
\cite{Pisarski:1999tv,Schmitt:2004et}, have different meaning than those used
in Sec. \ref{Sec:GL_with_DM}. Briefly, in case of fermion pairing, the terms
transverse/longitudinal correspond to the \emph{relative}-momentum structure of
the Cooper pair, while in case of the (vector) order parameter, they refer to
its dependence on the \emph{total} momentum of the Cooper pair. We believe that
no confusion can arise.}.

Second, we for simplicity disregard the gluonic fluctuations which would
otherwise make the superconducting phase transition first order
\cite{Noronha:2006cz}, and derive the GL free energy in the framework of the
Nambu--Jona-Lasinio (NJL) model \cite{Buballa:2003qv}. This is decently
justified by the fact that the coefficients of the GL free energy are to a
large extent universal, the only dynamics-dependent quantity being the critical
temperature \cite{Bailin:1984bm}. The fact that the NJL model does not capture
correctly the effects of the soft chromomagnetic gluons and thus predicts wrong
asymptotic behavior of the gap and critical temperature, should therefore not
matter as long as the critical temperature is appropriately adjusted.

Third, we neglect, as usual, the antiparticle degrees of freedom. As mentioned
above, in the ultrarelativistic limit this automatically projects out the
transverse part of the order parameter. Here we take the advantage of the fact
that the Dirac structure $\gamma_i$ can be achieved within a NJL-type model
with a contact, momentum-independent interaction. Finally, we neglect the
admixture of states with higher angular momentum \cite{Feng:2007bg,Feng:2008dh}.

Given the above assumptions, the mean-field approximation to the NJL model
amounts to a theory of noninteracting quasiquarks in the background of the
(possibly slowly-varying) order parameter $\phi_{ai}$, with the imaginary-time
propagator in the Nambu space, $\Psi\equiv(\psi,\psi^{\mathcal C})^T$, given by
\begin{equation}
\mathcal D^{-1}(i\omega_n,\vek k)=
\begin{pmatrix}
(i\omega_n-\xi_{\vek k})\gamma_0\Lambda^+_{\vek k} & \tilde\Phi\\
\Phi & (i\omega_n+\xi_{\vek k})\gamma_0\Lambda^-_{\vek k}
\end{pmatrix},
\label{quark_propagator}
\end{equation}
where we used the usual notation for $\epsilon_{\vek k}=\sqrt{\vek k^2+m^2}$
and the energy with respect to the Fermi level, $\xi_{\vek k}=\epsilon_{\vek
k}-\mu$. Also, $\Lambda^\pm_{\vek k}=\frac12\left[1\pm\frac1{\epsilon_{\vek
k}}\gamma_0(\vek\gamma\cdot\vek k+m)\right]$ are the standard positive/negative
energy state projectors, and
\begin{equation}
\Phi=\frac{\lambda^A_a}{\sqrt2}\gamma_i\phi_{ai},\quad
\tilde\Phi=\gamma_0\he\Phi\gamma_0,
\label{gap_matrix}
\end{equation}
where $(\lambda^A_a)_{bc}=-i\epsilon_{abc}$.

The GL free energy density at weak coupling and in the ultrarelativistic limit
then becomes
\begin{multline}
\mathcal F[\vec\phi_a]=\frac{7\zeta(3)p_{\text F}^3}{240\pi^4\mu
T_c^2}\left[2|\nabla_i\vec\phi_a|^2-|\vec\nabla\cdot\vec\phi_a|^2\right]\\
+\frac{\mu p_{\text F}}{3\pi^2}t|\vec\phi_a|^2+\frac{7\zeta(3)\mu p_{\text
F}}{480\pi^4T_c^2} (3\mathcal A+3\mathcal B-2\mathcal C),
\label{GL_micro_without}
\end{multline}
where $T_c$ is the critical temperature for the spin-one pairing. We denoted
$t=\frac{T}{T_c}-1$ and for the sake of an easy check with literature, kept
separate symbols for the Fermi momentum $p_{\text F}$ and the chemical
potential $\mu$, even though they actually coincide in the ultrarelativistic
limit. Derivation of the GL functional \eqref{GL_micro_without} is routine. Yet
we are not aware of any other independent calculation for spin-one color
superconductivity using the NJL model, and therefore provide some details in
Appendix \ref{App:GL_from_NJL}.

A comparison with Eq. \eqref{GL_without_DM} now yields the phenomenological
coefficients
\begin{gather*}
a_1=-2a_2=\frac{7\zeta(3)p_{\text F}^3}{120\pi^4\mu T_c^2},\quad
b=\frac{\mu p_{\text F}}{3\pi^2}t,\\
d_1=d_2=-\frac32d_3=\frac{7\zeta(3)\mu p_{\text F}}{160\pi^4T_c^2}.
\end{gather*}
The ratio of $d_2$ and $d_3$ is indicated in Fig.
\ref{Fig:phase_diagram_without} by a cross. In agreement with the full QCD
calculation we conclude that in the absence of the DM term, the weak-coupling
limit favors the CSL pattern.

In the presence of the DM term, the form of the ground state depends on the
parameter
$$
\xi=-\frac{c^2}{4a_1b}=-c^2\frac{90\pi^6T_c^2}{7\zeta(3)p_{\text F}^4t}.
$$
As remarked at the end of Sec. \ref{Subsec:complex_ferro}, the DM term slightly
increases the critical temperature. Eq. \eqref{Fgrsthelix} suggests that the
corrected critical temperature is given by $\xi=-1$ \footnote{In fact, this
quantitative estimate of the shift of the critical temperature should not be
really trusted, for in our derivation of the DM coefficient $c$, we neglect
analogous weak-interaction contributions to $b$ itself. However, the following
qualitative conclusions about the evolution of the ground state just below the
critical temperature remain valid.}, i.e.,
$$
t_c=c^2\frac{90\pi^6T_c^2}{7\zeta(3)p_{\text F}^4}.
$$
Just below the (new) critical temperature, $\xi$ is large and negative. With
decreasing temperature, it goes through a singularity to large positive values
and starts decreasing. (Of course, this ``singularity'' is completely
artificial and comes just from the definition of $\xi$.) In the tiny window
below the critical temperature and above the temperature at which $\xi=1/6$,
the ground state has therefore the planar structure, Eq. \eqref{planar:OP}.
Only for $\xi<1/6$ the system enters the distorted CSL phase---the axial phase,
the ground state being as in Eq. \eqref{axial:OP} with
$$
\alpha^2=\frac{1+4\xi}{3+2\xi},\quad
\beta^2=\frac{1-6\xi}{3+2\xi}.
$$
With further decreasing temperature, $\xi$ drops to zero and the ground
state relaxes to the CSL state. Needless to say that the temperature range in
which this evolution occurs is extremely narrow, but it is nevertheless
interesting to observe that the cooling of the system across the critical
temperature actually consists of a fast sequence of two phase transitions, both
of second order.

\subsection{DM term}
The last, and very important, missing ingredient in our analysis is the actual
value of the DM coefficient $c$. We will argue here that the parity-violating
DM term is naturally induced by the underlying weak interactions. Since the DM
term is bilinear in the order parameter, we will seek weak corrections to the
collective pairing mode propagator.

Before we begin the calculation, we make a remark about the discrete symmetries
of the DM term. As already stressed, it breaks the parity. At the same time, it
is invariant under charge conjugation (up to a total derivative). On the other
hand, it is well known that weak interactions in a single-fermion-family world
violate parity, but preserve the combined CP transformation. In the vacuum, the
DM term would therefore be prohibited, at least as a consequence of weak
interactions. However, charge conjugation is broken explicitly by the presence
of the dense medium, and the DM term therefore arises from the interplay of
weak interactions and many-body physics.

\begin{figure}
\includegraphics[scale=1]{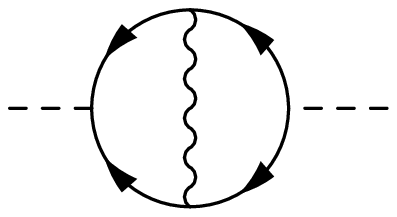}
\caption{Pairing field self-energy with $Z$-boson insertion.}
\label{Fig:Z_exchange}
\end{figure}
Taking into account Gaussian fluctuations above the mean-field Cooper pair
condensate, the pairing field propagator is given by a geometric series of the
fermion bubble diagrams \cite{Abuki:2006dv}. The most straightforward weak
correction is then the $Z$-boson exchange, depicted in Fig.
\ref{Fig:Z_exchange}. ($W^\pm$ bosons cannot be exchanged since we consider
just single-flavor quark matter.) Unfortunately, this does not work for the
following reason. As mentioned above, the dominant spin-one pairing pattern in
the ultrarelativistic limit is transverse in the relative momentum of the pair.
Roughly speaking, it corresponds to pairing with total momentum and orbital
momentum zero and total spin one, i.e., opposite helicity (and chirality). The
quark interaction vertex with the $Z$ boson can be parameterized as
$\gamma^\mu(u+v\gamma_5)$, where
\begin{align*}
u&=-\frac e{2\sin2\theta_{\text{W}}}\left(\frac53\sin^2\theta_{\text{W}}
-\cos^2\theta_{\text{W}}\right),\\
v&=-\frac e{2\sin2\theta_{\text{W}}},
\end{align*}
for $u$-type quarks, and
\begin{align*}
u&=\frac e{2\sin2\theta_{\text{W}}}\left(\frac13\sin^2\theta_{\text{W}}
-\cos^2\theta_{\text{W}}\right),\\
v&=\frac e{2\sin2\theta_{\text{W}}},
\end{align*}
for $d$-type quarks. Here $\theta_{\text{W}}$ is the Weinberg angle and $e$ is
the electric charge unit. The coupling of the $Z$ boson to two quarks of
opposite chirality in the loop in Fig. \ref{Fig:Z_exchange} then produces the
factor $(u+v)(u-v)=u^2-v^2$. However, to achieve parity breaking, one needs an
amplitude odd in $v$.

\begin{figure}
\includegraphics[scale=1]{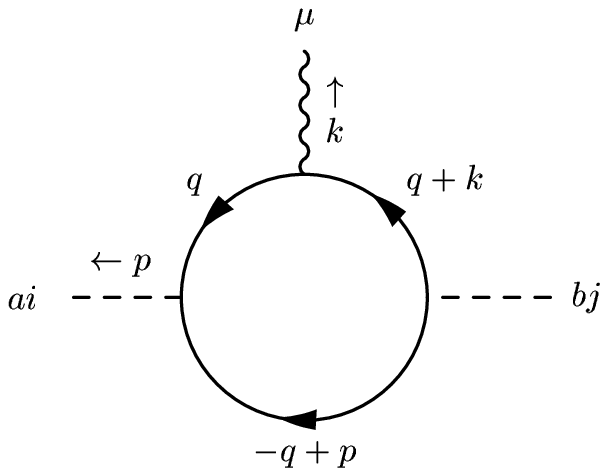}
\caption{Effective coupling of the pairing mode to the $Z$ boson. The labels of
the fermionic lines denote momentum flowing in the direction of the arrows.}
\label{Fig:Z_vertex}
\end{figure}
\begin{figure}
\includegraphics[scale=1]{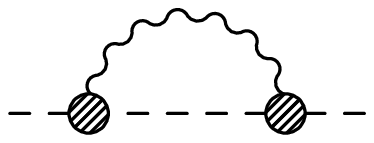}
\caption{Pairing field self-energy with $Z$-boson insertion. The hatched
circles denote the effective $Z$--diquark vertex defined by Fig.
\ref{Fig:Z_vertex}.}
\label{Fig:Z_loop}
\end{figure}

We therefore have to look for corrections of higher orders. Since each
propagator of a weak intermediate boson is suppressed by the huge weak scale of
the order of $100\,\text{GeV}$, it is most likely that any diagram with the
exchange of more than one heavy vector boson will be much smaller than a graph
with just one $Z$ boson, but including fluctuations of the pairing field.
Having in mind that the underlying interaction of the $Z$ boson with the weak
neutral quark current induces an effective coupling of $Z$ to the pairing field
$\vec\phi_a$ (see Fig. \ref{Fig:Z_vertex}), we thus anticipate that the leading
contribution to the DM term will be given by the graph in Fig.
\ref{Fig:Z_loop}. The fact that the diagram involves propagation of an
intermediate spin-one collective mode unfortunately means that we will be able
to provide only a rough, order-of-magnitude estimate of the DM coefficient $c$.

We first focus on the effective $Z$--diquark vertex, $\Gamma^\mu_{ai,bj}(k,p)$,
with the vector and axial-vector parts defined by
$$
\Gamma^\mu_{ai,bj}(k,p)=uV^\mu_{ai,bj}(k,p)+vA^\mu_{ai,bj}(k,p).
$$
For the sake of an analytical estimate of the DM term, we approximate these
vertex functions by their Taylor expansion at zero external momentum, working
again in the high-density approximation near the Fermi surface
\cite{Nardulli:2002ma}. For the vector part of the vertex we thus get
$$
V^\mu_{ai,bj}(0,0)=\frac{2\mu}{3\pi^2}\delta^{\mu0}\delta_{ab}\delta_{ij}
\log\frac{2\Lambda_{\text F}e^{\gamma-1}}{\pi T_c},
$$
where $\Lambda_{\text F}$ is an ultraviolet cutoff on the quark momentum; we
will choose it at the order of the Fermi momentum. On the contrary, the axial
part of the vertex is finite; it yields the expected parity-violating structure
and its Taylor series starts at the first order. Given the $g_{\mu\nu}$
structure of the $Z$-boson propagator, we just need to evaluate the axial vertex
for $\mu=0$,
$$
A^0_{ai,bj}(k,p)=\frac{7i\zeta(3)\mu p_{\text F}}{48\pi^4T_c^2}
\delta_{ab}\epsilon_{0ijk}(\vek k+2\vek p)_k.
$$

The DM term is produced by the diagram in Fig. \ref{Fig:Z_loop} where one of
the effective vertices is vector-like and the other one axial-like. Neglecting
the momentum-dependent part of the $Z$-boson propagator gives
\begin{multline*}
\frac1{M_Z^2}\sumint_k\mathcal
G_{ck,dl}(k)g_{\mu\nu}\left[A^\mu_{ai,ck}(k-p,p)V^\nu_{dl,bj}(-k+p,k)\right.\\
\left.+V^\mu_{ai,ck}(k-p,p)A^\nu_{dl,bj}(-k+p,k)\right],
\end{multline*}
where we used the shorthand notation for a sum-integral,
$$
\sumint_k=T\sum_n\int\frac{d^3\vek k}{(2\pi)^3},
$$
and $\mathcal G_{ai,bj}$ denotes the collective mode propagator. In the normal
phase, it is diagonal in the internal space, $\mathcal
G_{ai,bj}(k)\equiv\delta_{ab}\delta_{ij}\mathcal G(k)$. A comparison with the
DM term in Eq. \eqref{GL_with_DM} then results in an expression for the
coefficient $c$,
$$
c=\frac{7uv\zeta(3)\mu^2p_{\text F}}{36\pi^6T_c^2M_Z^2}
\left(\log\frac{2\Lambda_{\text F}e^{\gamma-1}}{\pi
T_c}\right)\sumint_k\mathcal G(k).
$$

The last sum-integral is apparently quadratically divergent. Part of the
divergence comes from the fact that we approximated the effective vertices in
Fig. \ref{Fig:Z_loop} with their low-momentum limits. We therefore make a rough
estimate by replacing the Matsubara sums with a frequency integral and putting
an ultraviolet cutoff, $\Lambda_{\text B}$, on the frequency--momentum
integration. Taking into account the appropriate volume measure and the fact
that the coefficient of the leading, $\mathcal O(\vek p^2)$, in the inverse
propagator of the collective mode is typically of order $10^{-1}$ in the
ultrarelativistic limit \cite{Brauner:2008td}, we find
\begin{equation}
c=\frac{7uv\zeta(3)\mu^2p_{\text F}}{36\pi^6T_c^2M_Z^2}\rho,\quad
\rho\approx10^{-1}\Lambda_{\text B}^2\log\frac{2\Lambda_{\text
F}e^{\gamma-1}}{\pi T_c}.
\label{c_final}
\end{equation}
The cutoff $\Lambda_{\text B}$ should be well above the characteristic scale of
the pair fluctuations in order not to suppress any physical contribution to the
integral; for the moment, we treat is as a free parameter.

\section{Phenomenological implications}
\label{Sec:phenomenology} With all coefficients of the GL functional we can now
readily determine the parameters of the ground state. In order to be able to
evaluate the condensation energy, we naively extrapolate the GL theory to zero
temperature. For a concrete calculation we consider the CSL structure with
$\bar d=d_1+\frac{d_2+d_3}3$, in the ultrarelativistic limit where $p_{\text
F}=\mu$; the orders of magnitude will nevertheless be the same for all phases.

First of all, the wavenumber of the helix in the ground state is given by
$$
k_{\text{DM}}=\frac{c}{2a_1}=\frac{5uv\mu\rho}{3\pi^2M_Z^2}.
$$
To appreciate the robustness of the helical ordering, we next compare the
condensation energies of the uniform condensation and of the helical ordering
itself. The first reads
$$
\mathcal F_{\Delta}=-\frac12b_0\sqrt{\mathcal A}=
-\frac{b_0^2}{4\bar d}=-\frac{4\mu^2T_c^2}{7\zeta(3)},
$$
where $b_0\equiv b/t$, while the latter is
$$
\mathcal
F_{\text{DM}}=-\frac{c^2}{4a_1}\sqrt{\mathcal A}=
-\frac{5\mu^4}{9\pi^6}\left(\frac{uv\rho}{M_Z^2}\right)^2.
$$
[For simplicity, we use the lower bound from Eq. \eqref{FminCSC} which is not
really saturated for the CSL state. However, this only changes the result by a
factor of order one.] Apparently, the energy gain from the formation of the
helical structure can be many orders of magnitude smaller than the condensation
energy of the superconducting state. It is therefore natural to ask whether the
long-range helical ordering will not be destroyed by thermal fluctuations of
the order parameter.

At low temperatures these will be dominated by the Nambu--Goldstone boson(s) of
the spontaneously broken symmetry. The energy density deposited in the thermal
fluctuations will therefore be that of the phonon gas with the phase velocity
$v_{\text{ph}}=1/\sqrt3$ (in the ultrarelativistic BCS limit), i.e.,
$3\pi^2T^4/90v_{\text{ph}}^3$. The helical ordering will be destroyed when this
becomes comparable to the condensation energy, $\mathcal F_{\text{DM}}$. This
leads to the characteristic temperature
$$
T_{\text{DM}}\sim\frac{\mu}{\pi^2}\left(\frac{uv\rho}{M_Z^2}\right)^{1/2}.
$$
[We dropped the unimportant numerical prefactor $(50/9\sqrt3)^{1/4}$ which is
close to one.]

For temperatures in the range $T_{\text{DM}}\lesssim T\lesssim T_c$, the system
will look just like the uniform spin-one color superconductor. Only at
temperatures lower than $T_{\text{DM}}$ the helical ordering will appear. This
change of behavior will, of course, not be associated with any phase transition.

In order to assess the importance of the DM effect, we now make a specific
numerical order-of-magnitude estimate. To that end, we need to know that the
electroweak couplings are such that $uv\sim10^{-2}$, the $Z$-boson mass is of
order $M_Z\sim100\,\text{GeV}$. We also choose the typical value of the
chemical potential to be $\mu=400\,\text{MeV}$ and set the fermionic cutoff
equally, $\Lambda_{\text F}=\mu$. There is some controversy in literature
regarding the size of the critical temperature, or gap, in spin-one color
superconductors. Let us be rather optimistic and assume that
$T_c\sim100\,\text{keV}$ \cite{Marhauser:2006hy}. Using Eq. \eqref{c_final} we
thus get
$$
\frac{k_{\text{DM}}}{\text{MeV}}\sim10^{-10}\left(\frac{\Lambda_{\text
B}}{\text{MeV}}\right)^2,\quad
T_{\text{DM}}\sim10^{-4}\Lambda_{\text B}.
$$

The choice of the bosonic cutoff $\Lambda_{\text B}$ is obviously crucial.
Our rough guess is slightly above the critical temperature itself,
$\Lambda_{\text B}\sim1\,\text{MeV}$. For this value the temperature scale at
which the helical ordering takes place becomes $0.1\,\text{keV}$, and the helix
wavelength is truly macroscopic---about a millimeter. The temperature scale
suggests that the phenomenon could take place in extremely cold quark matter
only (six orders of magnitude below the Fermi temperature) such as in very old
neutron stars.

In any case, however, the helical ordering in the ground state results in
peculiar low-energy properties of the system. As already remarked above, this
nonuniform ordering breaks both translational and rotational invariance,
leaving unbroken just their special combination. The system will therefore
exhibit anisotropic behavior. In helical ferromagnets, the Nambu--Goldstone
boson associated with the broken symmetry acts as a magnon when it propagates
transversely to the helix axis, and as a phonon when it propagates along
\cite{Kirkpatrick:2005ki}. The case of spin-one color superconductors will be
similar, in particular we expect that the Nambu--Goldstone mode will have
different phase velocities in the transverse and longitudinal directions. This
will in turn affect the thermodynamic properties of the system at very low
temperatures.

\section{Summary and discussion}
\label{Sec:conclusions} In this paper we have revised in detail the structure
of the ground state of spin-one color superconductors, composed of a single
quark flavor. In Sec. \ref{Sec:classification} we provided a complete
classification of possible ground states distinguished by the unbroken
continuous symmetry, assuming a local constant order parameter. In Sec.
\ref{Sec:GL_without_DM} we worked out the Ginzburg--Landau description of the
spin-one color superconductor near the critical temperature. We wrote down the
most general GL free energy up to fourth power in the order parameter and
determined its unique global minimum. We thus revealed that the noninert
$\varepsilon$ state is favored in part of the parameter space.

In Sections \ref{Sec:GL_with_DM} and \ref{Sec:microderivation} we then argued
that the ground state of a spin-one color superconductor will actually be
nonuniform as a result of a tiny parity-violating effect which is due to the
electroweak physics. Within the GL analysis, this effect is easily taken into
account by adding a new term to the free energy. In analogy with some
ferromagnetic materials, the ground state is found to exhibit helical ordering,
typically with a wavelength much longer than the characteristic scale of the
underlying many-fermion system. This leads in particular to an anisotropic
behavior of the system at low temperatures where the helical ordering sets.

In the course of our analysis, we made a number of simplifying assumptions
that we wish to discuss to some extent now. First, we assumed that the local
structure of the Cooper pair and hence also the classification of Sec.
\ref{Sec:classification} does not change when the DM term is switched on. This
is plausible when the wavelength of the helix is long enough, in particular
much longer than the Cooper pair size, so that the spatial modulation of the
condensate can be treated as a perturbation. In the realization of the
phenomenon considered in this paper, this is guaranteed by the huge ratio of the
electroweak vector boson mass to the pairing scale.

Second, while we considered just a positive-parity spin-one condensate,
electroweak interactions also induce a small admixture of a negative-parity
condensate which makes the GL analysis more involved \cite{Bailin:1984bm}. This
condensate does not interfere with the DM mechanism to the first order in the
electroweak effects, and we therefore neglect it here. Moreover, while this
condensate would be extremely difficult to detect, the helical ordering of the
ground state provides a macroscopic realization of parity violation.

Third, there are other sources of parity violation than direct weak-interaction
effects. For instance, condensation of pseudoscalar mesons due to finite
chemical potential also breaks parity \cite{Kryjevski:2004cw}. However, this is
most likely to happen only in the color-flavor-locked phase where all quarks
are paired, and hence it is not relevant for spin-one color superconductivity.

Fourth, the fact that the condensate is spatially nonuniform means that the
quarks pair with a small, yet nonzero total momentum. In this sense, the
helical state is analogous to the Larkin--Ovchinnikov--Fulde--Ferrell state
considered in crystalline color superconductors
\cite{Alford:2000ze,Casalbuoni:2003wh}. Different is, of course, the mechanism
of the inhomogeneous pairing: In crystalline color superconductors, it is
driven by a mismatch of the Fermi surfaces of the quarks to be paired, whereas
in the present case the Fermi momenta are exactly equal. The nonzero momentum
of the pair is induced by the DM term. The important difference between the
various phases considered here is that unlike the real axial, planar, and polar
states, the single-plane-wave states A and $\varepsilon$ carry nonzero (charge
as well as color) current. While in crystalline superconductors this is
balanced by a backflow of the unpaired fermions so that the total net current
in the ground state is zero \cite{Alford:2000ze}, the same issue in the A and
$\varepsilon$ helical phases yet remains to be clarified.

We observe that the axial and planar states which in a sense interpolate between
the CSL and polar phases, show up as physical ground states in the phase
diagram. This suggests that they should be seriously taken into consideration
even in the absence of the DM term. In particular, in Ref. \cite{Alford:2005yy}
the effect of color neutrality on spin-one color superconductivity as a
complement to the primary 2SC pairing was studied. It was shown that the polar
phase may be energetically preferred. Our results open the possibility of an
even more favorable state of the axial type, which can compensate the color
imbalance of the 2SC pairing and yet gain energy from condensation of all three
anticolors.

Finally, we emphasize that the helical ordering by the DM interaction is a
fairly general phenomenon; the only two prerequisites are a vector order
parameter and broken parity. There are several issues that deserve further
investigation. Besides the stress on the spin-one color superconductivity which
stems from the pairing of the other quark flavors, we would like to point out in
particular the peculiar properties of collective excitations. They will be
anisotropic even in the CSL-like axial state and may thus in principle provide a
clear manifestation of the nonuniform nature of the ground state. We are going
to study these issues in our future work.

\begin{acknowledgments}
I am obliged to two people for inspiration which was crucial for this work: to
Benedikt Binz, who told me about the fascinating phenomenon of helical
ferromagnetism, and to Ji\v{r}\'{\i} Ho\v{s}ek, for urging me to look for
weak-interaction effects in color superconductors. I am indebted to A. Schmitt
for sharing with me his insight in spin-one color superconductivity and for
numerous improving remarks and suggestions. I have also benefited from fruitful
discussions with H. Abuki, J. Ho\v{s}ek, D. H. Rischke, and Q. Wang. The
present research was supported by the Alexander von Humboldt Foundation.
\end{acknowledgments}

\emph{Note added}---After this work was completed, I learnt that the same phenomenon
was also studied in the context of heavy-fermion superconductors, see
\cite{Mineev:1994mi,Mineev:2008mi} and references therein. I thank K. Samokhin for
bringing this work to my attention.

\appendix
\section{Order parameter}
\label{App:OP} In this appendix we prove {\thname}s \ref{Thm:order_parameter} and
\ref{Thm:sep_invar}. The basic ingredient will be the polar decomposition, well known from
linear algebra.
\begin{theorem}[Polar decomposition]
\label{Thm:polar_decomp} Let $M$ be a square complex matrix. There is a unitary
matrix $U$ and a positive-semidefinite Hermitian matrix $H$ such that
\begin{equation}
M=UH
\label{polar_decomp}.
\end{equation}
The matrix $H$ is unique, while the matrix $U$ is unique if and only if $M$ is
nonsingular.
\end{theorem}
We will not prove this statement here but just remark that when $M$ is
singular, the matrix $U$ is determined up to a unitary transformation acting on
the eigenvectors of $M$ with zero eigenvalue.

\subsection{Proof of {\thname} \ref{Thm:order_parameter}}
Thanks to the polar decomposition \eqref{polar_decomp}, we can always make the order
parameter $\op$ Hermitian and positive-semidefinite by a suitable left unitary
transformation. We thus have $\op=S+iA$, where $S$ ($A$) is a real (anti)symmetric matrix.
As the next and last step we note that a diagonal orthogonal rotation \eqref{transfo_diag}
preserves (anti)symmetry, and thus transforms the matrices $S,A$ separately. The real
symmetric matrix $S$ can always be diagonalized by such a transformation, which brings the
order parameter to the form \eqref{op_special}.

\subsection{Proof of {\thname} \ref{Thm:sep_invar}}
The diagonal orthogonal transformation \eqref{transfo_diag} preserves
Hermiticity and the spectrum, hence Eq. \eqref{op_sep_invar} says $U\op'=\op$,
where $\op'=R^T\op R$ and both $\op$ and $\op'$ are Hermitian and
positive-semidefinite. However, {\thname} \ref{Thm:polar_decomp} asserts that
the Hermitian part of the polar decomposition is unique. Therefore, we
necessarily find $\op=\op'=R^T\op R$ and $U\op=\op$, as was to be proved.

\section{Inequalities between quartic invariants}
\label{App:inequalities} In this appendix we prove {\thname}s \ref{Thm:veta1},
\ref{Thm:veta2}, \ref{Thm:veta3}, and \ref{Thm:veta5}. Several proofs will be
based on the well known Cauchy inequality which asserts that for any two
complex vectors $u,v$, we have
\begin{equation}
\Bigl|\sum_iu_i^*v_i\Bigr|^2\leq\Bigl(\sum_i|u_i|^2\Bigr)\Bigl(\sum_i|v_i|^2\Bigr).
\label{cauchy}
\end{equation}
The inequality in Eq. \eqref{cauchy} is saturated if and only if the vectors
$u,v$ are collinear.

\subsection{Proof of {\thname} \ref{Thm:veta1}}
Let us denote the (real and positive) eigenvalues of $\op\he\op$ as
$\delta_i^2$. Then $\mathcal A=\left(\sum_i\delta_i^2\right)^2$ and $\mathcal
B=\sum_i\delta_i^4$. The second inequality in Eq. \eqref{veta1} as well as the
condition for its saturation follow immediately. For the first inequality, set
$u_i=\delta_i^2$ and $v_i=1$. The Cauchy inequality \eqref{cauchy} then gives
$\mathcal A\leq3\mathcal B$, as was to be proved. This inequality is saturated
when all the eigenvalues $\delta_i^2$ are equal, that is, $\op\he\op$ is
proportional to the unit matrix. The order parameter $\op$ is then unitary up
to a real scale factor and can be brought to the CSL form by a left unitary
transformation.

\subsection{Proof of {\thname} \ref{Thm:veta2}}
\label{App_sub:proof_veta2} Using the complex-vector notation, $\mathcal
C=|\vec\phi_a\cdot\vec\phi_b|^2$ so that we obviously have $\mathcal C\geq0$
with the equality when all scalar products $\vec\phi_a\cdot\vec\phi_b$ are
zero. Writing $\vec\phi_a$ in terms of its real and imaginary parts, this first
of all requires $|\mathrm{Re}\,\vec\phi_a|=|\mathrm{Im}\,\vec\phi_a|$ and
$\mathrm{Re}\,\vec\phi_a\perp\mathrm{Im}\,\vec\phi_a$ for all rows $\vec\phi_a$
of the matrix $\op$. One of the vectors, say $\vec\phi_1$, can then be cast
into a special form by a right orthogonal rotation, so that the order parameter
becomes
$$
\op=\begin{pmatrix}
z_1 & iz_1 & 0\\
z_2 & z_3 & z_4\\
z_5 & z_6 & z_7
\end{pmatrix}.
$$
Let us assume that $z_1\neq0$ (if not, we go on processing the second row in
the same manner). The orthogonality, $\vec\phi_1\cdot\vec\phi_2=0$, then
implies $z_2+iz_3=0$, and from $\vec\phi_2\cdot\vec\phi_2=0$, $z_4=0$ follows.
The vectors $\vec\phi_1$ and $\vec\phi_2$ are therefore collinear so that
$\vec\phi_1$ can be made zero by a suitable left unitary transformation.
Proceeding in the same manner, we next nullify $\vec\phi_2$ and end up with
just the third row of $\op$, being proportional to $(1,i,0)$. This is
equivalent to the A-phase order parameter.

For the second inequality in Eq. \eqref{veta2}, recall that Hermitian matrices
span a real vector space with a scalar product defined as $(A,B)=\tr(AB)$. If
we set $Z=\he\op\op$, we have $\mathcal B=\tr(ZZ)=||Z||^2$ and $\mathcal
C=\tr(Z^*Z)=(Z^*,Z)$. Therefore, we get
$$
0\leq||Z-Z^*||^2=||Z||^2+||Z^*||^2-2(Z,Z^*)=2(\mathcal B-\mathcal C).
$$
We thus prove $\mathcal C\leq\mathcal B$ with the equality if and only if
$Z=Z^*$. This translates into the requirement that $\Delta_{ai}^*\Delta_{aj}$
is real, i.e., all scalar products of columns of $\op$ must be real. However,
the order parameter may always be transformed by a left unitary matrix to the
form
$$
\op=\begin{pmatrix}
a_1 & z_1 & z_2\\
0 & a_2 & z_3\\
0 & 0 & a_3
\end{pmatrix},
$$
with real $a_i$ and complex $z_i$. Applying step by step the above reality requirement we
find that the whole matrix actually has to be real.

\subsection{Proof of {\thname} \ref{Thm:veta3}}
As already mentioned in Appendix \ref{App_sub:proof_veta2}, we can define a
scalar product of two Hermitian matrices by the trace of their matrix product.
We now introduce some further notation. We first define an orthonormal basis,
$T_a,a=0,\dotsc,8$, so that $(T_a,T_b)=\delta_{ab}$. These matrices are simply
defined as $T_0=\openone/\sqrt3$ and $T_a=\lambda_a/\sqrt2,a=1,\dotsc,8$, in
terms of the standard Gell-Mann matrices. Expanding a given Hermitian matrix in
this basis, e.g. $A=a_aT_a$, the scalar product becomes $(A,B)=a_ab_a$.

As above we denote $Z=\he\op\op$. Recalling that the basis matrices $T_2,T_5,T_7$ are
imaginary while all others are real, we can see that the complex conjugation in $Z^*$ just
changes the sign of the coordinates $z_2,z_5,z_7$. Introducing finally the shorthand
notation,
\begin{equation}
\begin{split}
u^2&=z_0^2,\quad w^2=z_2^2+z_5^2+z_7^2,\\
v^2&=z_1^2+z_3^2+z_4^2+z_6^2+z_8^2,
\end{split}
\label{uvw}
\end{equation}
we obtain the expressions for the three invariants,
\begin{align*}
\mathcal A&=(\tr Z)^2=(Z,\openone)^2=3u^2,\\
\mathcal B&=(Z,Z)=u^2+v^2+w^2,\\
\mathcal C&=(Z^*,Z)=u^2+v^2-w^2.
\end{align*}
This implies $\mathcal B+\mathcal C=2(u^2+v^2)$, whence we immediately get the
desired inequality \eqref{veta3}. It will be saturated if and only if $v=0$,
i.e., if the real part of $Z\equiv\he\op\op$ will be proportional to the unit
matrix. After substitution for the order parameter from \eqref{op_special}, it
is straightforward to show that this is equivalent to the condition stated in
{\thname} \ref{Thm:veta3}.

\subsection{Proof of {\thname} \ref{Thm:veta5}}
This inequality is most tricky because it does not hold for all matrices $\op$
but just for those satisfying $\mathcal C\leq\frac19\mathcal A$. We will need
another auxiliary claim.
\begin{theorem}
\label{Thm:vwaux} Every Hermitian positive-semidefinite matrix $Z$ satisfies the
following inequality,
\begin{equation}
v\geq w\sqrt3-u\sqrt2,
\label{vwaux}
\end{equation}
where the non-negative quantities $u,v,w$ are defined by Eq. \eqref{uvw}. The
inequality is saturated if and only if $Z$ is of the $\varepsilon$ type.
\end{theorem}
Proof.---By a suitable diagonal orthogonal rotation, $Z\to R^TZR$, we can
always make the coordinates $z_5$ and $z_7$ vanish so that $z_2=\pm w$. Let us
without lack of generality assume that $z_2=w$. Pick up a test vector as the
eigenvector of $\lambda_2$ with the eigenvalue $-1$,
$|t\rangle=\frac1{\sqrt2}(1,-i,0)^T$. The expectation value of $Z$ is
$$
\langle t|Z|t\rangle=\frac{u}{\sqrt3}+\frac{z_8}{\sqrt6}-\frac{w}{\sqrt2}.
$$
Positive-semidefiniteness requires that this be non-negative, which leads to
$$
w\sqrt3\leq z_8+u\sqrt2\leq v+u\sqrt2.
$$
The inequality \eqref{vwaux} is thus proved. In order for it to be saturated,
we must have $v=z_8$, i.e., $z_1=z_3=z_4=z_6=0$, and $z_8=w\sqrt3-u\sqrt2$. The
matrix $Z$ then reads
$Z=\frac{u}{\sqrt3}\openone+\frac{w}{\sqrt2}\lambda_2+\frac1{\sqrt2}
(w\sqrt3-u\sqrt2)\lambda_8$, which has the $\varepsilon$ form with
$\Delta_1=w/\sqrt2$ and $\Delta_2=u\sqrt3-w\sqrt2$.

We now get to the proof of {\thname} \ref{Thm:veta5}. For $\mathcal
C\leq\frac19\mathcal A$ we find using {\thname} \ref{Thm:veta3},
$$
\mathcal B-\mathcal C\geq\frac23\mathcal A-2\mathcal C\geq\frac23\mathcal
A-\frac29\mathcal A=\frac49\mathcal A,
$$
whence $\frac2{\sqrt3}u=\frac23\sqrt{\mathcal A}\leq\sqrt{\mathcal B-\mathcal
C}=w\sqrt2$. This means that the right-hand side of the inequality
\eqref{vwaux} is non-negative, and the inequality can be equivalently squared.
This yields
\begin{gather*}
v^2\geq(w\sqrt3-u\sqrt2)^2=3w^2-2\sqrt6uw+2u^2,\\
\mathcal C=u^2+v^2-w^2\geq2w^2-2\sqrt6uw+3u^2=(w\sqrt2-u\sqrt3)^2,\\
\sqrt{\mathcal C}\geq u\sqrt3-w\sqrt2=\sqrt{\mathcal A}-\sqrt{\mathcal
B-\mathcal C}.
\end{gather*}
The saturation condition is the same as that for the inequality \eqref{vwaux}.
Since the order parameter $\op$ is Hermitian and positive-semidefinite without
lack of generality, it is given by the unique square root of $Z=\he\op\op$. It
is therefore $\varepsilon$-like if and only if $Z$ is $\varepsilon$-like.

\section{GL free energy from NJL model}
\label{App:GL_from_NJL} In the NJL model, the mean-field free energy of a
superconductor is determined by the quasifermionic excitations above the Fermi
sea, and is given in terms of the fermion propagator \eqref{quark_propagator}
as
$$
\mathcal F=-\frac{T}{2\Omega}\tr\log\mathcal D^{-1}.
$$
Here the trace is taken in the functional sense and the factor $\frac12$ comes
from doubling of the number of degrees of freedom in the Nambu formalism. The
condensate contribution, quadratic in the order parameter, is not included here
for it essentially only serves to adjust the GL coefficient $b$ to zero at the
critical temperature. Writing the inverse propagator as usual as $\mathcal
D^{-1}=\mathcal D_0^{-1}+\Sigma$, where the self-energy $\Sigma$ is
off-diagonal in the Nambu space and contains the pairing field, expanding in
powers of the order parameter up to fourth order, and subtracting the free
energy of the normal phase, the GL functional becomes
$$
\mathcal F=\frac{T}{2\Omega}\left[ \frac12\tr(\mathcal
D_0\Sigma)^2+\frac14\tr(\mathcal D_0\Sigma)^4\right].
$$

The traces here lead to momentum integrals which are evaluated in the
high-density approximation \cite{Nardulli:2002ma}: The integral over the
momentum magnitude, $|\vek k|$, is replaced with one over the energy measured
with respect to the Fermi sea, $\xi_{\vek k}$, multiplied by the density of
states at the Fermi surface, $\mathcal N=\frac{\mu p_{\text F}}{2\pi^2}$.

Assuming for a moment the generic form of the gap matrix,
$\Phi(x)=\phi_a(x)T_a$, where $T_a$ is a set of momentum-independent matrices
in the Dirac and fermion-species (flavor) space, the individual terms (referred
to as the ``gradient'', ``mass'', and ``quartic'' terms with obvious meaning)
in the GL free energy density become
\begin{multline*}
\text{gradient term}=\frac{7\zeta(3)p_{\text F}^2}{32\pi^2\mu^2T_c^2}\mathcal
N\sum_{\vek
q}\vek q^2\varphi^*_{a,\vek q}\varphi_{b,-\vek q}\\
\times\left\langle(\hat{\vek p}\cdot\hat{\vek q})^2
\tr(\Lambda^+_{\vek p}\he T_a\Lambda^-_{-\vek p}T_b)\right\rangle_{\vek p},
\end{multline*}
$$
\text{mass term}=\frac12\mathcal N t\phi^*_a\phi_b\left\langle
\tr(\Lambda^+_{\vek p}\he T_a\Lambda^-_{-\vek p}T_b)\right\rangle_{\vek p},
$$
\begin{multline*}
\text{quartic
term}=\frac{7\zeta(3)}{32\pi^2T_c^2}\mathcal N\phi^*_a\phi_b\phi^*_c\phi_d\\
\times\left\langle\tr(\Lambda^+_{\vek p}\he T_a\Lambda^-_{-\vek p}
T_b\Lambda^+_{\vek p}\he T_c\Lambda^-_{-\vek p}T_d)\right\rangle_{\vek p}.
\end{multline*}
In all the expressions, angular brackets denote averaging over directions of
the indicated momentum, and the hats unit vectors. Also, $\varphi_{a,\vek q}$
are the Fourier components of the order parameter, defined as in Eq.
\eqref{OP_FT}, and the traces are now taken in the Dirac and flavor space. The
above expressions are valid for zero as well as finite fermion mass and it is
understood that in the projectors $\Lambda^\pm_{\pm\vek p}$ the energy
$\epsilon_{\vek p}$ and momentum $\vek p$ are replaced with their values on the
Fermi level, i.e., $\mu$ and $p_{\text F}\hat{\vek p}$.

One can readily check the validity of the general result above on a
particularly simple example, namely the (relativistic) BCS superconductor. Here
we have just a single order parameter, $\phi$, and the corresponding matrix
$T=\gamma_5$, which ensures positive-parity pairing. All the Dirac traces are
then equal to two and the only nontrivial angular average is $\langle(\hat{\vek
p}\cdot\hat{\vek q})^2\rangle_{\vek p}=\frac13$, so that we recover the
well-known GL functional of the BCS theory \cite{Bailin:1984bm},
$$
\mathcal F_{\text{BCS}}=\frac{7\zeta(3)p_{\text F}^3}{96\pi^4\mu
T_c^2}|\vec\nabla\phi|^2 +\frac{\mu p_{\text F}}{2\pi^2}t|\phi|^2
+\frac{7\zeta(3)\mu p_{\text F}}{32\pi^4T_c^2}|\phi|^4.
$$

\subsection{GL functional for spin-one color superconductors}
The calculation of the GL coefficients may be pushed forward by making a
particular assumption on the structure of the gap matrix. It is instructive to
divide the calculation in two steps and assume first the spin-one pairing
structure $\Phi=\phi_{ai}Q_a\gamma_i$, where the $Q_a$'s are yet unspecified
matrices in the flavor space, normalized for the sake of convenience by
$\tr_{\text F}(Q_a\he Q_b)=\delta_{ab}$. The bilinear terms in the GL free
energy are then straightforward to evaluate using the explicit form of the
energy projectors,
\begin{multline}
[\vek\gamma]:\quad\text{gradient term}=\frac{7\zeta(3)p_{\text F}^3}{96\pi^4\mu T_c^2}\\
\times\left[\left(1-\frac{p_{\text F}^2}{5\mu^2}\right)|\nabla_i\vec\phi_a|^2-
\frac{2p_{\text F}^2}{5\mu^2}|\vec\nabla\cdot\vec\phi_a|^2\right],
\label{g:grad}
\end{multline}
\begin{equation}
[\vek\gamma]:\quad\text{mass term}= \frac{\mu p_{\text
F}}{2\pi^2}\left(1-\frac{p_{\text F}^2}{3\mu^2}\right)t|\vec\phi_a|^2.
\label{g:mass}
\end{equation}

To calculate the Dirac trace in the quartic term, we resort to the
ultrarelativistic limit. In this case, the energy projectors reduce the spatial
$\gamma$-matrices to their transverse parts,
$$
\gamma_{\perp i}=\mathcal P_{ij}\gamma_j,\quad \mathcal P_{ij}=\delta_{ij}-\hat
p_i\hat p_j,
$$
and the Dirac trace simplifies to
$$
\frac12\tr_{\text D}(\gamma_{\perp i}\gamma_{\perp j}
\gamma_{\perp k}\gamma_{\perp l})=
2(\mathcal P_{ij}\mathcal P_{kl}-\mathcal P_{ik}\mathcal P_{jl}+
\mathcal P_{il}\mathcal P_{jk}).
$$
Using the identity
$$
\langle\mathcal P_{ij}\mathcal P_{kl}\rangle_{\vek p}=
\frac1{15}
(6\delta_{ij}\delta_{kl}+\delta_{ik}\delta_{jl}+\delta_{il}\delta_{jk}),
$$
we finally get
\begin{multline}
[\vek\gamma]:\quad\text{quartic term}=\frac{7\zeta(3)\mu p_{\text
F}}{240\pi^4T_c^2}
(3\delta_{ij}\delta_{kl}-2\delta_{ik}\delta_{jl}+3\delta_{il}\delta_{jk})\\
\times\phi^*_{ai}\phi_{bj}\phi^*_{ck}\phi_{dl} \tr_{\text F}(\he Q_aQ_b\he
Q_cQ_d).
\label{g:quart}
\end{multline}
Eqs. \eqref{g:grad}, \eqref{g:mass}, and \eqref{g:quart} summarize the general
Ginzburg--Landau functional for a pairing with the Dirac structure $\gamma_i$.
In particular for the spin-one pairing considered in this paper, the gap matrix
has the form \eqref{gap_matrix}. The flavor trace is then $\tr_{\text F}(\he
Q_a Q_b\he Q_cQ_d)=\frac14(\delta_{ab}\delta_{cd}+\delta_{ad}\delta_{bc})$.
Putting all the pieces together, we arrive at the expression for the free
energy in Eq. \eqref{GL_micro_without}.

While the transverse pairing can be achieved with the Dirac structure
$\gamma_i$, setting $\Phi=\phi_{ai}Q_a\gamma_0\gamma_i$ leads (in the
ultrarelativistic limit) to a purely longitudinal pairing. In this case, the
Dirac traces become trivial and we just quote the final result,
\begin{multline}
[\gamma_0\vek\gamma]:\quad\text{gradient term}=\frac{7\zeta(3)p_{\text F}^3}{96\pi^4\mu T_c^2}\\
\times\left[\left(1-\frac{4p_{\text F}^2}{5\mu^2}\right)|\nabla_i\vec\phi_a|^2+
\frac{2p_{\text F}^2}{5\mu^2}|\vec\nabla\cdot\vec\phi_a|^2\right],
\label{gg0:grad}
\end{multline}
\begin{equation}
[\gamma_0\vek\gamma]:\quad\text{mass term}= \frac{\mu p_{\text
F}}{2\pi^2}\left(1-\frac{2p_{\text F}^2}{3\mu^2}\right)t|\vec\phi_a|^2.
\label{gg0:mass}
\end{equation}
\begin{multline}
[\gamma_0\vek\gamma]:\quad\text{quartic term}=\frac{7\zeta(3)\mu p_{\text
F}}{480\pi^4T_c^2}
(\delta_{ij}\delta_{kl}+\delta_{ik}\delta_{jl}+\delta_{il}\delta_{jk})\\
\times\phi^*_{ai}\phi_{bj}\phi^*_{ck}\phi_{dl} \tr_{\text F}(\he Q_aQ_b\he
Q_cQ_d).
\label{gg0:quart}
\end{multline}
As in the previous case, Eqs. \eqref{gg0:grad} and \eqref{gg0:mass} are valid
for arbitrary fermion mass, while the quartic term \eqref{gg0:quart} was for
simplicity derived in the ultrarelativistic limit.

In the end we would like to remark that a simple special case of our GL
functional for spin-one color superconductors is the pairing of quarks of a
single color and two flavors in the antisymmetric flavor-singlet channel,
studied by Buballa \emph{et al.} \cite{Buballa:2002wy}. In this case, the flavor
matrix is given by $Q=\frac1{\sqrt2}\tau_2\otimes\mathcal P^{(c)}_3$, where
$\tau_2$ is the Pauli matrix in flavor space and $\mathcal P^{(c)}_3$ the
projector on the third quark color. The order parameter is a complex vector,
$\vec\phi$, and one immediately finds, in the ultrarelativistic limit,
\begin{multline*}
\mathcal F_{\vek\gamma}=\frac{7\zeta(3)p_{\text F}^3}{240\pi^4\mu T_c^2}
\left[2|\nabla_i\vec\phi|^2-|\vec\nabla\cdot\vec\phi|^2\right]
+\frac{\mu p_{\text F}}{3\pi^2}t|\vec\phi|^2\\
+\frac{7\zeta(3)\mu p_{\text
F}}{240\pi^4T_c^2}\left[3(\he{\vec\phi}\cdot\vec\phi)^2-
|\vec\phi\cdot\vec\phi|^2\right],
\end{multline*}
\begin{multline*}
\mathcal F_{\gamma_0\vek\gamma}=\frac{7\zeta(3)p_{\text F}^3}{480\pi^4\mu
T_c^2} \left[|\nabla_i\vec\phi|^2+2|\vec\nabla\cdot\vec\phi|^2\right]
+\frac{\mu p_{\text F}}{6\pi^2}t|\vec\phi|^2\\
+\frac{7\zeta(3)\mu p_{\text
F}}{960\pi^4T_c^2}\left[2(\he{\vec\phi}\cdot\vec\phi)^2+
|\vec\phi\cdot\vec\phi|^2\right].
\end{multline*}
It is amusing to observe that the transverse and longitudinal cases differ in
the sign of the $|\vec\phi\cdot\vec\phi|^2$ term, which results in
qualitatively different forms of the ground state. In the $\vek\gamma$ case the
negative sign leads to a polar-like state, $\vec\phi\sim(0,0,1)^T$. On the
other hand, the positive sign in the $\gamma_0\vek\gamma$ case (actually
considered in \cite{Buballa:2002wy}) leads to an A-like state,
$\vec\phi\sim(1,i,0)^T$, with peculiar properties such as the existence of a
single Nambu--Goldstone boson with a quadratic dispersion relation---the spin
wave \cite{Buballa:2002wy,Brauner:2005di}.

\end{document}